\documentclass[a4paper,11pt]{article}
\pdfoutput=1 

\usepackage{jcappub}
\usepackage{graphicx}
\usepackage{dcolumn}
\usepackage{amssymb,amsmath,bm,bbold}
\usepackage{color}
\usepackage[dvipsnames,table]{xcolor}
\usepackage{xfrac}
\usepackage{aas_macros}
\usepackage{mathrsfs}
\usepackage{subcaption}
\usepackage{rotating}
\usepackage{chngcntr}
\usepackage{lipsum}
\usepackage{units}
\usepackage{multirow}
\usepackage{xspace}
\usepackage{xfrac}
\usepackage{mathtools}
\usepackage{comment}
\usepackage{pifont}
\usepackage{soul}
\newcommand{\cmark}{\ding{51}}%
\newcommand{\xmark}{\ding{55}}%
\newcommand{\nv}{\hat{\bf n}}
\newcommand{\disp}{\boldsymbol{\Psi}}
\newcommand{\colore}{{\tt CoLoRe}}

\usepackage[T1]{fontenc} 
\usepackage{natbib}
\title{{\tt CoLoRe}: fast cosmological realisations over large volumes with multiple tracers}
\author[a]{C\'esar Ram\'irez-P\'erez,}
\author[b]{Javier Sanchez,}
\author[c]{David Alonso,}
\author[a,d]{Andreu Font-Ribera}

\affiliation[a]{Institut de F\'isica d'Altes Energies. The Barcelona Institute of Science and Technology, Campus UAB, 08193 Bellaterra (Barcelona), Spain}
\affiliation[b]{Fermi National Accelerator Laboratory, PO Box 500, Batavia, IL 60510, USA}
\affiliation[c]{Department of Physics, University of Oxford, Denys Wilkinson Building, Keble Road, Oxford OX1 3RH, UK}
\affiliation[d]{Department of Physics and Astronomy, University College London, Gower Street, London WC1E 6BT, UK}

\emailAdd{cramirez@ifae.es}

\abstract{
We present \colore, a public software package to efficiently generate synthetic realisations of multiple cosmological surveys.
\colore\ can simulate the growth of structure with different degrees of accuracy, with the current implementation supporting lognormal fields, first, and second order Lagrangian perturbation theory.
\colore\ simulates the density field on an all-sky light-cone up to a desired maximum redshift, and uses it to generate multiple 2D and 3D maps: galaxy positions and velocities, lensing (shear, magnification, convergence), integrated Sachs-Wolfe effect, line intensity mapping, and line of sight skewers for simulations of the Lyman-$\alpha$ forest. We test the accuracy of the simulated maps against analytical theoretical predictions, and showcase its performance with a multi-survey simulation including DESI galaxies and quasars, LSST galaxies and lensing, and SKA intensity mapping and radio galaxies. We expect \colore\ to be particularly useful in studies aiming to characterise the impact of systematics in multi-experiment analyses, quantify the covariance between different datasets, and test cross-correlation pipelines for near-future surveys.}

\begin{document}

\vspace*{\headsep}\vspace*{\headheight}
\begin{footnotesize}
\hfill FERMILAB-PUB-21-562-PPD\\
\end{footnotesize}

\maketitle
\flushbottom

\section{Introduction}
\label{sec:intro}

Ongoing and future cosmological surveys will explore large volumes with multiple tracers to study dark energy, inflation and massive neutrinos. Spectroscopic surveys such as the Dark Energy Spectroscopic Instrument (DESI, \cite{DESI_whitepaper}), \textit{Euclid} \cite{2011arXiv1110.3193L} and \textit{Roman} \cite{2015arXiv150303757S} will collect tens of millions of precise galaxy redshifts. Meanwhile, the Vera Rubin Observatory's Legacy Survey of Space and Time (LSST, \cite{2009arXiv0912.0201L}) will photometrically observe billions of galaxies and provide an exquisite weak lensing map over a large fraction of the sky. Lensing maps will also be obtained from future experiments observing the cosmic microwave background (CMB), such as the Simons Observatory (SO, \cite{2019JCAP...02..056A}) and CMB-S4 \cite{2016arXiv161002743A}. Finally, large catalogs of radio galaxies and 21cm intensity maps will be provided by experiments such as the Square Kilometer Array (SKA, \cite{2020PASA...37....7S}) or HIRAX \citep{2016SPIE.9906E..5XN}.

In order to obtain robust cosmological constraints from these large and complex datasets, it is important to be able to efficiently generate \textit{mocks}, synthetic realisations of the data.
For instance, mocks are often used to compare survey strategies, to test the analysis pipeline, to study the impact of astrophysical or instrumental contaminants, and to estimate the covariance of the measurements.

When generating mocks one needs to trade off realism for computing costs.
It is just not feasible to generate hundreds or thousands of N-body simulations covering volumes of tens of cubic gigaparsecs, let alone hydrodynamic simulations that could model baryonic effects.
On the other side of the spectrum, lognormal realisations \citep{1991MNRAS.248....1C} offer an efficient way of obtaining simplified mock catalogs with the correct distribution only on large, linear scales (see for instance \cite{2016MNRAS.459.3693X,2017JCAP...10..003A,2021JCAP...03..095M}).
Lagrangian Perturbation Theory (LPT, \citep{2002PhR...367....1B}) has inspired several approximated methods that can reproduce the distribution of matter on intermediate, mildly non-linear scales.
These include \texttt{PTHalos} \citep{2002MNRAS.329..629S,2013MNRAS.428.1036M,2015MNRAS.447..437M}, \texttt{Pinocchio} \citep{2002MNRAS.333..623T}, \texttt{COLA} \citep{2013JCAP...06..036T}, \texttt{QPM} \cite{2014MNRAS.437.2594W}, \texttt{PATCHY} \citep{2014MNRAS.439L..21K} \texttt{ICE-COLA} \citep{aizard_icecola}, \texttt{L-PICOLA} \citep{Howlett_picola}, \texttt{HALOGEN} \citep{savila_halogen_2015} and \texttt{EZMocks} \citep{2015MNRAS.446.2621C}.

The best cosmological inference will come from joint analyses of multiple cosmological probes, each providing independent and complementary information. Two of the most important challenges in these joint analyses will be characterising the effects of systematics affecting several experiments, and estimating the cross-covariance between the different 2D and 3D datasets, with partially overlapping area and redshift ranges. This publication addresses the need of simultaneously simulating these surveys in a coherent and efficient framework.

We present \colore\ (Cosmological Lofty Realization), a parallel C code for generating fast mock realizations of multiple cosmological surveys\footnote{The code is publicly available at \url{https://github.com/damonge/CoLoRe}.}. \colore\ can simulate the growth of structure using either a lognormal model or LPT (at 1st or 2nd order), and it can simulate a plethora of cosmological tracers: photometric and spectroscopic galaxies, weak lensing, intensity mapping, Integrated Sachs-Wolfe effect and CMB lensing, or the Lyman-$\alpha$ forest in the spectra of high-redshift quasars. 
It has been designed in a highly modular fashion, making it easy to add new tracers or more complex models of growth of structure. 
It uses both OpenMP and MPI parallelisation, and it is specially suited to run with multiple nodes in high performance computing facilities. 

The paper is organised as follows. In Section \ref{sec:meth} we describe in detail \colore{}, its code structure, the cosmological assumptions made, and the list of tracers already available.
The validation of \colore\ to generate reliable mocks for intensity mapping and for Lyman-$\alpha$ forest studies was already presented in previous work \cite{2014MNRAS.444.3183A, 1912.02763}. 
In Section \ref{sec:res} we validate its ability to simulate galaxy clustering and weak lensing statistics, and discuss the computing and memory requirements to run large simulations.
Finally, in Section \ref{sec:conclusions} we draw the conclusions.

\section{Methods}\label{sec:meth}
  \subsection{Overall code structure}\label{ssec:meth.code}
    \colore~ is written in a modular way that makes it relatively straightforward to modify (e.g. to add a new non-linear structure formation model, or a new tracer of the density fluctuations). In a standard run, \colore~ goes through the steps listed below, each of which is associated with a compartmentalised piece of code:
    \begin{enumerate}
      \item {\bf Initialisation.} \colore{} interprets the configuration file, allocates the resources needed to carry out the requested simulation, and initialises a number of cosmological quantities (redshift-distance relation, growth history, linear matter power spectrum, background densities of all source tracers).
      \item {\bf Predictions:} \colore{} produces theoretical predictions for the three-dimensional power spectrum of all biased matter tracers in the lognormal approximation. This is mostly useful when using the lognormal structure formation model.
      \item {\bf Gaussian random fields.} Two three-dimensional Cartesian grids are generated containing the linear matter overdensity $\delta^L_M({\bf x})$ and the Newtonian gravitational potential $\phi_N({\bf x})$ at redshift $z=0$. The grid is sufficiently large to hold a sphere of comoving radius $\chi(z_{\rm max})$, where $z_{\rm max}$ is the maximum redshift of the run. The spatial resolution of the simulation is set by $N_{\rm grid}$, the number of grid cells into which the box is divided in each dimension. The grid cell size is therefore approximately $\Delta x=L_{\rm box}/N_{\rm grid}\simeq 2\chi(z_{\rm max})/N_{\rm grid}$.
      \item {\bf Physical density field.} The Gaussian overdensity $\delta^L_M({\bf x})$ is transformed into a non-linear, physical overdensity field $\delta_M({\bf x})$ through one of the structure formation models supported by \colore. This is done in the lightcone (i.e. the value of the field at comoving position ${\bf x}$ is $\delta_M(t(|{\bf x}|),{\bf x})$, where $t(\chi)$ is the cosmic time at comoving distance $\chi$), with the observer located at the center of the Cartesian box. The physical density field is such that $\delta\geq-1$ everywhere. The gravitational potential is also evolved in the lightcone assuming linear growth.
      \item {\bf Density normalisation.} \colore{} uses non-linear transformations to generate biased tracers of the matter overdensity. In general, these can be written as
      \begin{equation}\label{eq:bias}
        1+\delta_k = \frac{B_k(\delta_M)}{\langle B_k(\delta_M)\rangle},
      \end{equation}
      where $B_k$ is the non-linear biasing relation for tracer $k$. Due to the non-linearity of these relations, the ergodic average in the denominator of the previous equation is not necessarily equal to 1 (even if $\langle\delta_M\rangle=0$), and therefore the normalising factor ensures that $\langle\delta_k\rangle=0$ for all biased tracers. Since densities are defined on the lightone, the normalisation factors are computed at this stage independently for several redshift shells by averaging over grid cells.
      \item {\bf Get Cartesian information.} At this stage the overdensity and Newtonian potential grids are distributed across computer nodes as slabs of equal width $N_{\rm slab} =N_{\rm grid}/N_{\rm nodes}$. 
      Before proceeding further, \colore{} collects all information available in these slabs and needed by each of the tracers requested for this simulation. This involves any data product not involving line-of-sight integrals or interpolations, which are dealt with in later stages. For instance, this is when source catalogs are generated by Poisson-sampling the biased density field. All tracers are endowed with a method {\tt tracer\_set\_cartesian} that collects this information.
      \item {\bf Redistribution into beams.} In order to carry out line-of-sight integrals and interpolations, \colore{} redistributes tracer data so each node has access to all the data in a set of sky regions, labelled ``beams'', covering the full range of redshifts $0\leq z\leq z_{\rm max}$. Each beam is defined using the {\tt HEALPix} pixellation scheme \citep{2005ApJ...622..759G}, as the region of the celestial sphere covered by a given low resolution pixel. The {\tt HEALPix} $N_{\rm side}$ resolution parameter used to define these beams is chosen to be large enough that the full dataset is approximately evenly split between computer nodes. All tracers have an associated method {\tt tracer\_distribute} in charge of distributing the tracer data in each node's slab to all other nodes whose beams intersect with it. Note that, although the tracer information is now distributed across nodes through beams, the density and Newtonian potential grids are still distributed in slabs.
      \item {\bf Get beam information.} Any calculation involving a line-of-sight integral (e.g. gravitational lensing) or interpolation (e.g. Lyman-$\alpha$ skewers) is done after the tracers have been redistributed into beams. The calculation is done in three stages:
      \begin{enumerate}
          \item {\sl Preprocessing.} Initialisation of any necessary quantities (e.g. setting all variables that eventually hold a gravitational lensing calculation to zero). Each tracer has an associated function {\tt tracer\_beams\_preproc} in charge of doing this.
          \item {\sl Loop through slabs.} All nodes send their current slab of the density and Newtonian potential grids to the node on their right (assuming periodic boundary conditions). Once the new slab is received, each node gathers the necessary information from it (e.g. the contribution to a lensing convergence integral from the section of the Newtonian potential held in that slab) and adds it to each tracer\footnote{Note that all quantities calculated at this stage (e.g. integrals and interpolations) are linear and additive on $\delta_M$ and $\phi_N$.}. \colore{} carries out this calculation through a method {\tt tracer\_get\_beam\_properties} associated with each tracer. This is repeated $N_{\rm nodes}$ times, at which point all nodes have had access to the full density and potential grids.
          \item {\sl Post-processing.} Each tracer finishes off any calculation still needed after having gathered all information in the preceding step (e.g. multiplying maps by an overall normalization factor). This is done by a method {\tt tracer\_beams\_postproc} associated with each tracer.
      \end{enumerate}
      \item {\bf Write output.} Each tracer writes all its information (e.g. in the form of maps or catalogs) to file through a method {\tt write\_tracer}. \colore{} uses the {\tt FITS} standard in most cases, although it is also possible to save source catalog data as {\tt ASCII} or {\tt HDF5} files.
    \end{enumerate}

    Modifying \colore{} to support a new structure formation model would involve implementing it as part of step 4 above, with no effect on the rest of the code. Adding a new type of tracer would involve creating the corresponding tracer methods for it enumerated above ({\tt \_set\_cartesian}, {\tt \_distribute}, {\tt \_beams\_preproc}, {\tt \_get\_beam\_properties}, {\tt \_beams\_postproc}, and {\tt write\_}).

    The assumptions made by \colore{} to compute the background cosmological quantities (step 1) are described in Section \ref{ssec:meth.cosmo}. Section \ref{ssec:meth.box} describes the Gaussian density fields and the different non-linear structure formation models supported by \colore{} (steps 3 and 4). Section \ref{ssec:meth.tr} describes in detail the calculations carried out for each of the tracers, including the bias models supported (steps 5-8). Finally, the theory predictions computed by \colore{} for lognormal fields (step 2) are discussed in Section \ref{ssec:meth.lognormal}.

  \subsection{Cosmological assumptions}\label{ssec:meth.cosmo}
    When generating simulated observations, \colore{} makes a number of assumptions about the underlying cosmological model. We describe these here.

    \colore{} assumes a flat $w$CDM cosmological background, characterised, at low redshifts, by 3 cosmological parameters: the background matter density $\Omega_M$, the current expansion rate $H_0$, and a constant dark energy equation of state parameter $w$. The expansion rate is thus given by
    \begin{equation}
      H(z)=H_0\,\left[\Omega_M(1+z)^{3}+(1-\Omega_M)(1+z)^{3(1+w)}\right],
    \end{equation}
    in terms of which the comoving distance is\footnote{Note that we use units with $c=1$ throughout.}
    \begin{equation}
       \chi(z)=\int_0^z\frac{dz'}{H(z')}.
    \end{equation}

    Matter density perturbations are governed by a linear matter power spectrum at $z=0$, $P_0(k)$, which must be provided to \colore{} on input, and is then normalised to the chosen value of $\sigma_8$. If the power spectrum is needed on scales larger than those provided, it is extrapolated assuming a power-law behaviour $P_0(k)\propto k^{n_s}$ on small $k$, where $n_s$ is the scalar spectral index (also provided on input).

    Finally, \colore{} assumes a self-similar growth for the linear matter overdensity: $\delta_M^L({\bf x},z)=\delta_M^L({\bf x},0)D(z)$, where $D(z)$ is the linear growth factor. $D(z)$ is calculated from the cosmological parameters by solving the differential equation
    \begin{equation}\label{eq:growth1}
      \frac{d}{da}\left(a^3\,H(a)\frac{dD}{da}\right)=\frac{3}{2}\Omega_M(a)\,aH(a)\,D(a),
    \end{equation}
    where $a=1/(1+z)$ is the scale factor.

    Although internally \colore{} uses ``$h$-inverse'' units (i.e. distances are given in units of ${\rm Mpc}\,h^{-1}$), all simulation outputs involve observable quantities (redshift and angles), and therefore are insensitive to this choice.

  \subsection{Matter box}\label{ssec:meth.box}
    The first step after initialising the cosmological model in a standard \colore{} run is the generation of a Gaussian realisation of the linear matter inhomogeneities $\delta_M^L$ at $z=0$ on a Cartesian cubic grid. This is done by drawing the Fourier coefficients of $\delta_M^L$ as independent Gaussian random numbers from the input linear matter power spectrum using the Box-Muller transform with variance:
    \begin{equation}
      \sigma^2({\bf k})=\frac{P_0(k)}{(\Delta k)^3},
    \end{equation}
    where $\Delta k \equiv 2\pi/L_{\rm box}$ is the sampling rate in Fourier space. \colore{} can alternatively apply a Gaussian smoothing kernel with scale $R_G$ to the linear power spectrum when generating the linear Fourier coefficients. This may be useful to control the behaviour of the non-linear overdensity field (see discussion in Section \ref{ssec:meth.lognormal}).

    At the same time, \colore{} populates a similar cartesian grid with the values of the Newtonian gravitational potential $\phi_N({\bf x})$, related to the matter inhomogeneities in Fourier space via:
    \begin{equation}
      \phi_N({\bf k})=-\frac{3}{2}H_0^2\,\Omega_M\,\frac{\delta_M^L({\bf k})}{k^2}.
    \end{equation}

    The linear matter overdensity thus generated is then transformed into a physical (i.e. positive-definite) non-linear matter overdensity in the lightcone using one of the three structure formation models currently supported by \colore{}, which we describe below.

    \subsubsection{Lognormal fields}\label{sssec:meth.box.ln}
      Lognormal fields were first proposed and analysed by \cite{1991MNRAS.248....1C} as a possible way to describe the distribution of matter in the Universe. A lognormal random field $x_{\rm LN}$ is defined in terms of a Gaussian random field $x_{\rm G}$ through the local transformation
      \begin{equation}\label{eq:ln1}
        x_{\rm LN}=\exp x_{\rm G}.
      \end{equation}
      One of the nice properties of these fields is that, while the Gaussian variable $x_{\rm G}$ is allowed to take any values in $(-\infty,+\infty)$, $x_{\rm LN}$ can only take positive values by construction. Furthermore, as discussed in \cite{1991MNRAS.248....1C}, the density field evolved along Lagrangian trajectories according to the linear velocity field along can be well described by a lognormal distribution, which justifies the use of lognormal fields from a physical point of view. In order to obtain a lognormal overdensity field with zero mean from a Gaussian field, the transformation (\ref{eq:ln1}) must be slightly varied as follows:
      \begin{equation}\label{eq:ln2}
        1+\delta_{\rm LN}=\exp\left(\delta_G - \frac{\sigma_G^2}{2}\right),
      \end{equation}
      where $\sigma_G$ is the variance of the Gaussian overdensity field.

      Lognormal density fields have been used in the past by different collaborations
      in order to perform fast galaxy mock realisations \cite{2005MNRAS.362..505C,
      2011MNRAS.416.3017B,2011MNRAS.415.2892B,2012JCAP...01..001F,2011A&A...534A.135L}, and are, therefore, a well established tool. Since the lognormal transformation is a simple, local modification of the density field, it is by far the fastest and most memory-efficient structure formation model implemented in \colore{}.

      However, the simplicity of the lognormal transformation implies that lognormal fields cannot be expected to describe all higher-order correlators of the density field (e.g. bispectra), to give rise to filamentary structure, or to reproduce the small-scale properties of the density field correctly \citep{2010MNRAS.403..589K}, and therefore this kind of mock realisations have a limited applicability.

      Within this framework, the non-linear overdensity is generated in \colore{} by applying Eq. \ref{eq:ln2} to the linear overdensity field evolved in the lightcone assuming linear growth (which is applied to both $\delta_G$ and $\sigma_G$ in this equation).

      Caution must be exercised when making use of lognormal fields. Since the lognormal transformation involves the exponentiation of a Gaussian field, large $(\gg1)$ values of $\delta_G$ lead to much larger fluctuations in $\delta_{\rm LN}$. Thus, if the amplitude of $\delta_G$, characterised by its standard deviation $\sigma_G$, is large, the resulting lognormal field will exhibit a large degree of inhomogeneity, with an enormous variance dominated by extreme fluctuations in a small number of voxels. This behaviour can be avoided through the use of the Gaussian smoothing kernel described above. This modifies the linear and non-linear power spectra in an analytically predictable manner.

      Note that other common implementations of the lognormal transformation to generate mock cosmological realisations (e.g. \cite{2011MNRAS.416.3017B}) have employed a different method to avoid this problem. Instead of using the input power spectrum to generate $\delta_G$ and then transform it into $\delta_{\rm LN}$, the input power spectrum is taken to be that of the final $\delta_{\rm LN}$. The inverse lognormal transformation is thus applied to the input power spectrum at the level of the two-point correlator (see Section \ref{ssec:meth.lognormal}), to obtain the power spectrum with which the Gaussian field is generated. \colore{} does not explicitly support this method, since it runs contrary to the idea of treating the lognormal transformation as a non-linear structure formation model. It would be, however, possible to pass as input to \colore{} the ``Gaussianised'' power spectrum in order to obtain the desired matter power spectrum at $z=0$ in the non-linear matter fluctuations.

    \subsubsection{Lagrangian perturbation theory}\label{sssec:meth.box.lpt}
      Lagrangian perturbation theory (LPT) \citep{2002PhR...367....1B} provides an alternative fast method to generate non-linear physical matter overdensities, which has been used in the past to generate mock galaxy catalogs \cite{2013MNRAS.428.1036M,2015MNRAS.446.2621C,2015MNRAS.447..437M}. The beauty of LPT lies in its ability to capture non-linear aspects of the Eulerian matter overdensity (e.g. the formation of filamentary structure) by carrying out low-order perturbation theory calculations in the displacement field. The basic premise is as follows: the linear overdensity is used to predict the Lagrangian displacement of a set of massive test particles with respect to their original unperturbed positions. The non-linear density field is then given by the density of the displaced test particles which is, by definition, positive-definite. \colore{} supports the generation of Lagrangian displacements at first and second order in perturbation theory. These first and second-order displacements at $z=0$ are scaled with the corresponding growth factors before interpolating the test particles onto a grid. We provide a brief overview of LPT here, and direct the reader to \cite{2002PhR...367....1B} for further details.

       Let ${\bf x}(t)=a(t)\,\left[{\bf q}+\disp({\bf q},t)\right]$ be the physical position of a particle starting at comoving coordinates ${\bf q}$. $\disp({\bf  q},t)$ is the so-called Lagrangian displacement vector. In the Newtonian approximation, the motion of these particles is governed by Newton's second law, which is sourced by the gravitational potential caused by the particles themselves. This leads to two coupled equations that can be summarized into a single equation for the divergence of $\disp$:
       \begin{equation}
         J\left({\sf J}^{-1}\nabla\right)\cdot\left(\disp''+a\,H\,\disp'\right)=\frac{3}{2}a^2H^2\,\Omega_M(J-1),
       \end{equation}
       where all derivatives are taken with respect to conformal time $d\tau\equiv dt/a$, ${\sf J}_{ij}\equiv\delta_{ij}+\partial_i\Psi_j$ is the Jacobian of the Lagrangian flow, and $J\equiv{\rm det}({\sf J})$. Note that the matter overdensity is given by $1+\delta=J^{-1}$.

       At second order in the displacement field, and discarding all curl-like components of $\disp$, the solution is given by
       \begin{equation}
         \disp({\bf q},a)=D(a)\,\nabla\varphi_{\rm LPT}^{(1)}({\bf q})+D^{(2)}(a)\,\nabla\varphi_{\rm LPT}^{(2)}({\bf q}).
       \end{equation}
       Here $D(a)$ is the linear growth factor, satisfying Eq. \ref{eq:growth1}, $D^{(2)}$ is the second-order growth factor, satisfying
       \begin{equation}\label{eq:growth2}
         \frac{d}{da}\left(a^3\,H(a)\frac{dD^{(2)}}{da}\right)=\frac{3}{2}\Omega_M(a)\,aH(a)\,\left(D^{(2)}(a)-[D(a)]^2\right),
       \end{equation}
       and $\varphi_{\rm LPT}^{(1,2)}$ are the first- and second-order LPT potentials, given by
       \begin{align}\label{eq:lpt_pois1}
         &\nabla^2\varphi_{\rm LPT}^{(1)}=-\delta_{\rm M}^L,\\\label{eq:lpt_pois2}
         &\nabla^2\varphi_{\rm LPT}^{(2)}=\frac{1}{2}\sum_{ij}\left[\partial_i^2\varphi_{\rm LPT}^{(1)}\partial_j^2\varphi_{\rm LPT}^{(1)}-\left(\partial_i\partial_j\varphi_{\rm LPT}^{(1)}\right)^2\right].
       \end{align}
       \colore{} solves the LPT Poisson equations (\ref{eq:lpt_pois1} and \ref{eq:lpt_pois2}) in Fourier space using the Gaussian overdensity field as input, and computes the second-order growth factor using the approximation \citep{2002PhR...367....1B}
       \begin{equation}
         D_2(a)=-\frac{3}{7}\left[D(a)\right]^2\left[\Omega_M(a)\right]^{-1/143}.
       \end{equation}
       Once the displacement vector has been calculated and applied to a set of test particles initially located at the centers of the Cartesian grid cells, the density field is calculated by interpolating the displaced positions onto the grid\footnote{To do this, \colore{} supports three standard mass-assignment methods: nearest-grid-point (NGP), cloud-in-cell (CIC), and triangular-shaped cloud (TSC) \citep{1981csup.book.....H}.}.

       LPT is able to generate a more realistic non-linear density field than the lognormal model at the cost of significantly higher memory requirements and longer computation times. While generating the lognormal overdensity does not require additional resources beyond those used to generate the $\delta_M^L$ and $\phi_N$ grids, generating the first-order displacement requires three additional Cartesian grids to hold the components of $\disp$, and second-order LPT demands an additional 5 grids to store the Hessian of the first-order LPT potential (the array holding the Gaussian density field can be reused to store one of the 6 independent components of the Hessian). The number of fast Fourier transforms needed, which dominate the total computation time, is also different in each case: none in the case of the lognormal model, 4 in the case of first-order LPT, and 13 for second-order LPT.

  \subsection{Tracers}\label{ssec:meth.tr}
    \colore{} is able to generate simulated observations for a variety of cosmological tracers of the same underlying matter fluctuations. The details of the calculations involved in each of the supported tracer types, carried out in steps 5 to 8 of the procedure outlined in Section \ref{ssec:meth.code}, is discussed in detail here.

    \subsubsection{Projected maps}\label{sssec:meth.tr.proj}
      \begin{figure}
          \centering
          \includegraphics[width=0.7\textwidth]{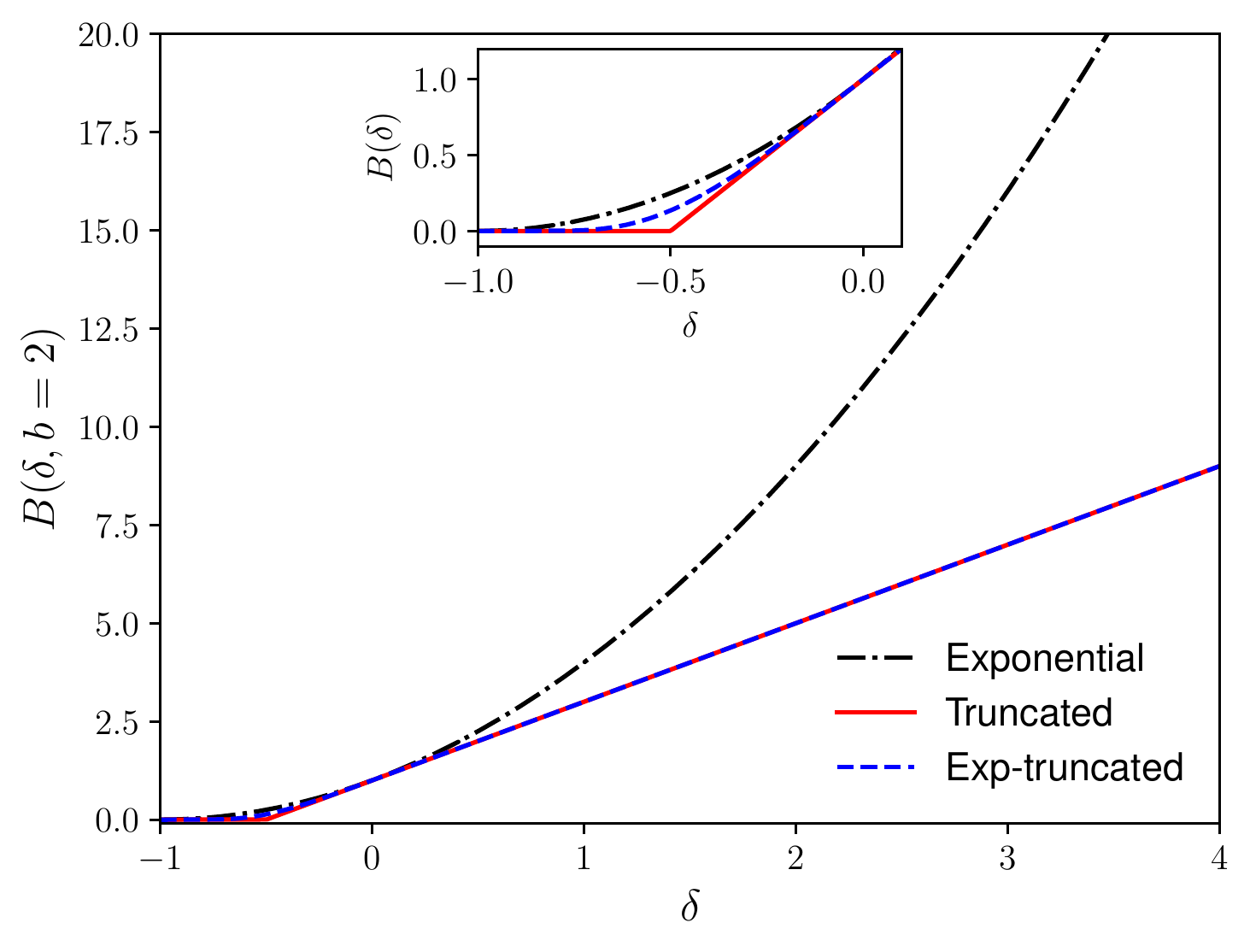}
          \caption{Bias models implemented in \colore{}. The exponential model preserves the ``lognormality'' of the field if using the lognormal structure formation model, but it can lead to numerically unstable results in the presence of sufficiently large fluctuations in the Gaussian field. The exp-truncated model can be used to curb this behaviour.}
          \label{fig:bias}
      \end{figure}
      \begin{figure}
          \centering
          \includegraphics[width=0.8\textwidth]{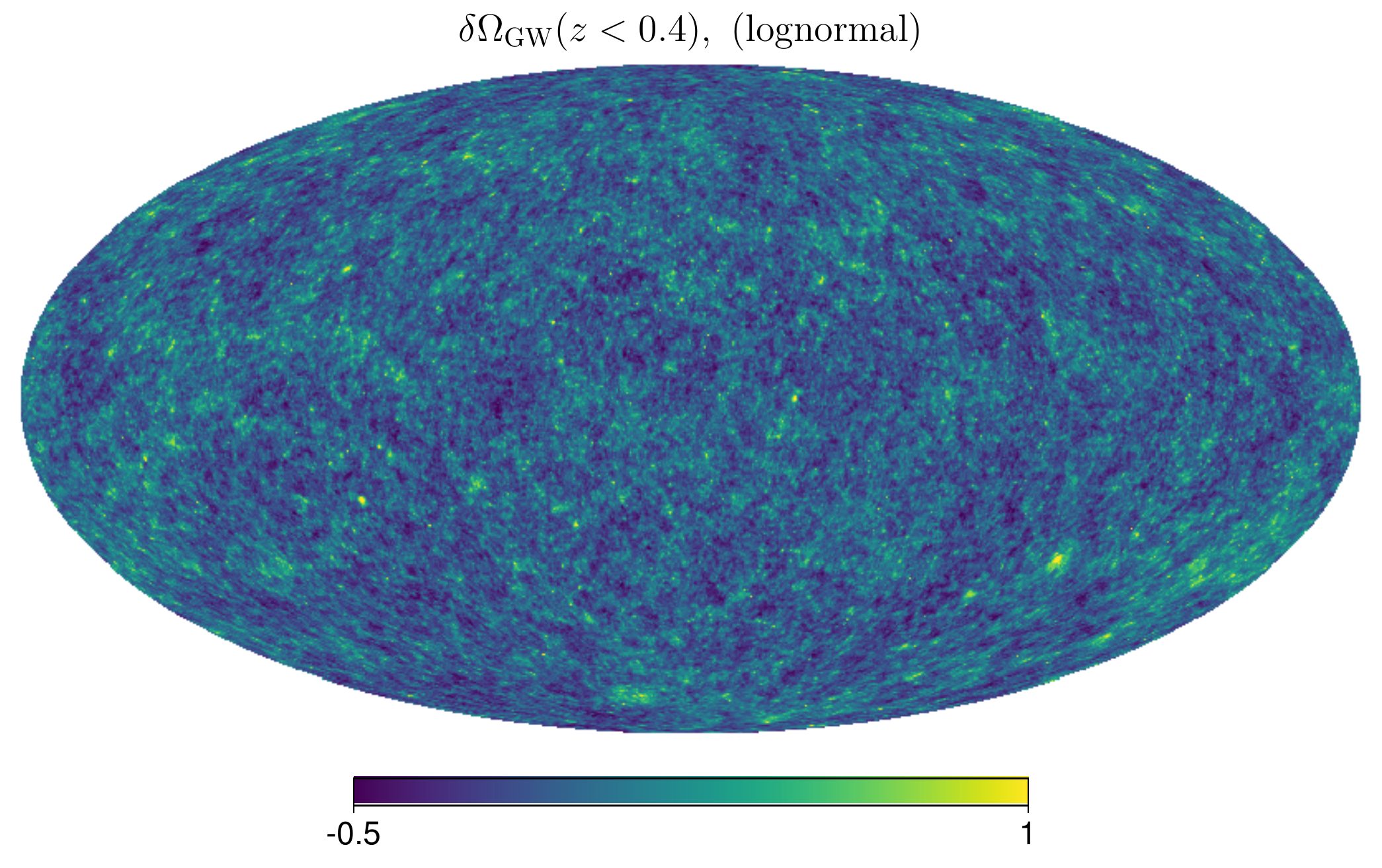}
          \includegraphics[width=0.8\textwidth]{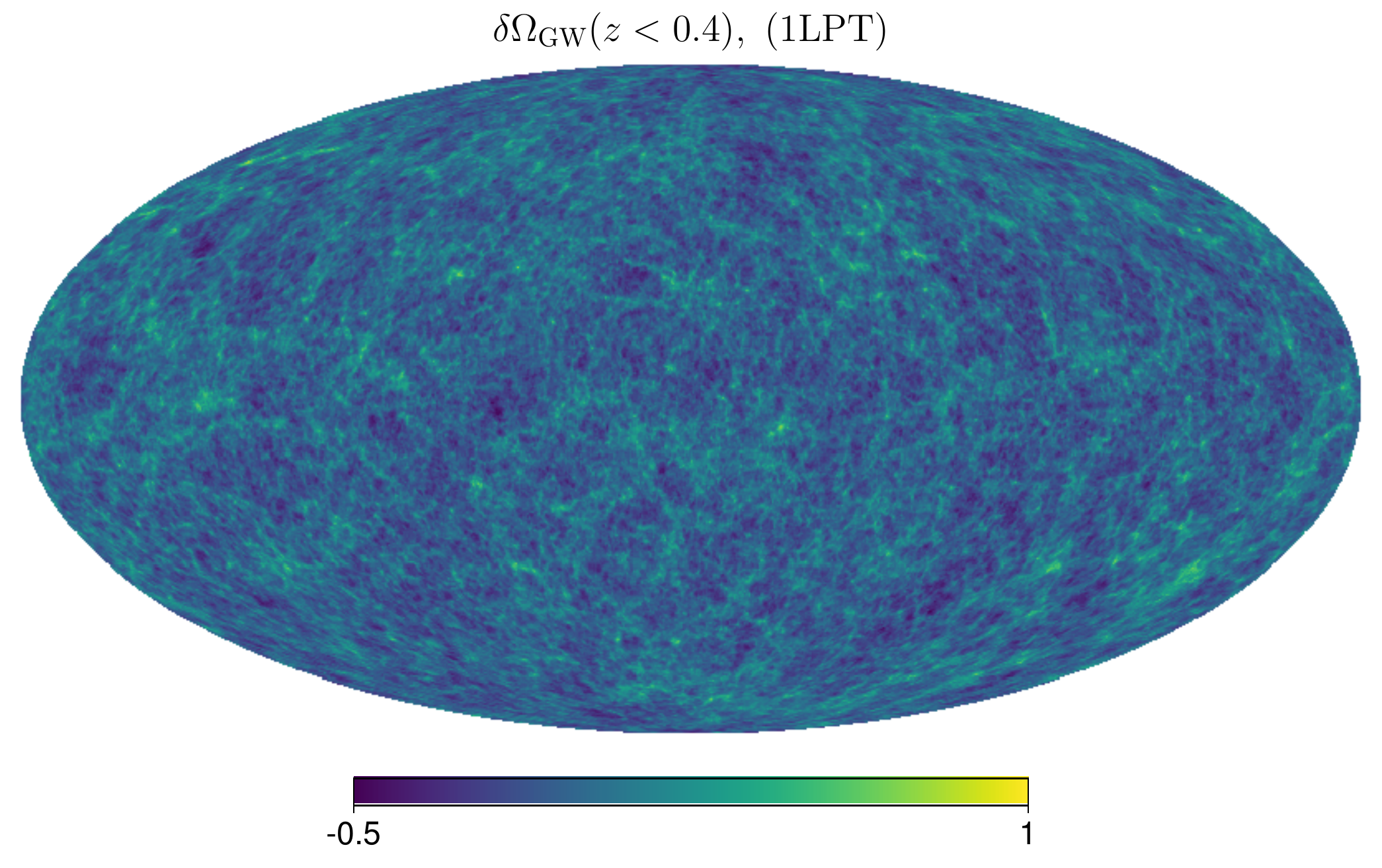}
          \caption{Simulated maps of the anisotropic stochastic gravitational wave background from astrophysical sources at redshifts $z<0.4$ using the models of \cite{2020MNRAS.493L...1C}. The top and bottom plots show simulations using the lognormal and first-order LPT sructure formation models respectively. The former is characterised by strong positive fluctuations on a few regions, while the latter displays the more physical filamentary structure of the cosmic web.}
          \label{fig:gws}
      \end{figure}
      The simplest tracer supported by \colore{} is the so-called ``custom projected tracer''. The associated observable is the overdensity in a biased tracer of the matter fluctuations projected onto the celestial sphere after integrating over an arbitrary radial kernel:
      \begin{equation}\label{eq:proj}
        \Delta_W(\nv)=\int dz\,W(z)\,\delta_W(z,\chi(z)\nv),
      \end{equation}
      where the integral is over redshift $z$, $W(z)$ is the tracer's radial kernel, and $\delta_W$ are the three-dimensional fluctuations in the tracer, related to the matter fluctuations via Eq. \ref{eq:bias}.

      \colore{} supports the following three local bias models, although more can be easily added to the code:
      \begin{align}\label{eq:bias_1}
          &\text{\bf Exponential:} && B_k(\delta_M)=(1+\delta_M)^{b_k},\\\label{eq:bias_2}
          &\text{\bf Truncated:} && B_k(\delta_M)= {\rm Max}(1+b_k\delta_M,0),\\\label{eq:bias_3}
          &\text{\bf Exp-truncated:} && B_k(\delta_M)=\left\{
          \begin{array}{cc}
            \exp\left[b_k\delta_M/(1+\delta_M)\right] & \delta_M\leq0 \\
              1+b_k\delta_M& \delta_M > 0
          \end{array}
          \right..
      \end{align}
      These three models are shown in Figure \ref{fig:bias} for $b_k=2$, and are designed to be positive definite ($B_k(\delta_M)>0$ for $\delta_M\in[-1,\infty)$), and to reduce to a linear biasing relation ($B_k(\delta_M)\simeq1+b_k\delta_M$) for small fluctuations. Given the modular nature of \colore, different biasing models could be easily added by the user.

      The integral in Eq. \ref{eq:proj} is calculated as follows: for each sky pixel, an imaginary line of sight connecting it with the observer at the box center is subdivided into intervals of constant comoving distance $\Delta\chi$, commensurate with the Cartesian cell size $\Delta x$. The value of $\delta_M$ at the center of each interval is calculated from the Cartesian grid using trilinear interpolation, and is then translated into the corresponding $\delta_W$. The integral is then calculated as a sum over all intervals along the line of sight.

      Custom projected tracers can be used in \colore{} to make simulated maps of a wide variety of two-dimensional cosmological anisotropic observables that correlate with the large-scale structure. As an example, Figure \ref{fig:gws} shows simulated maps of the anisotropic stochastic gravitational wave background from astrophysical sources at $z<0.4$ according to the models of \cite{2020MNRAS.493L...1C}. The figure shows results for the lognormal and first-order LPT structure formation models, showcasing the morphological differences between them.

    \subsubsection{Integrated Sachs-Wolfe effect}\label{sssec:meth.tr.isw}
      The time evolution in the gravitational potential at late times due to the accelerated background expansion causes an energy loss or gain in a background of photons which correlates with the large-scale structure. This is the so-called integrated Sachs-Wolfe effect (ISW, \citep{1967ApJ...147...73S}).

      The fluctuation in the temperature of a background of photons with black-body spectrum emitted at redshift $z_*$ is given by
      \begin{equation}
        \left.\frac{\Delta T}{T}\right|_{\rm ISW}(\nv)=2\int_0^{z_*}dz \frac{\dot{\phi}_N(z,\chi(z)\nv)}{(1+z)\,H(z)}.
      \end{equation}
      Assuming linear growth, appropriate on the large scales on which the ISW is relevant, one can approximate $\dot{\phi}_N(z)=H(z)[f(z)-1]\phi_N(z)$,
      where $f\equiv d\log D/d\log a$ is the growth rate.

      The ISW tracer is therefore equivalent to the custom projected tracer described in the previous section with the Newtonian potential $\phi_N$ taking the role of $\delta_W$, and with a kernel
      \begin{equation}
        W_{\rm ISW}(z)=\frac{f(z)-1}{1+z}\Theta(z<z_*),
      \end{equation}
      where $\Theta$ is the Heaviside function. Thus, the same numerical methods described in Section \ref{sssec:meth.tr.proj} are used by \colore{} to generate simulated ISW maps. The top panel of Figure \ref{fig:kisw} shows an example of a simulated map of the ISW effect for $z_*=0.5$, characterised by features on very large scales due to the $1/k^2$ relation between gravitational potential and matter overdensity.

    \subsubsection{Gravitational lensing}\label{sssec:meth.tr.gl}
      \begin{figure}
        \centering
        \includegraphics[width=0.8\textwidth]{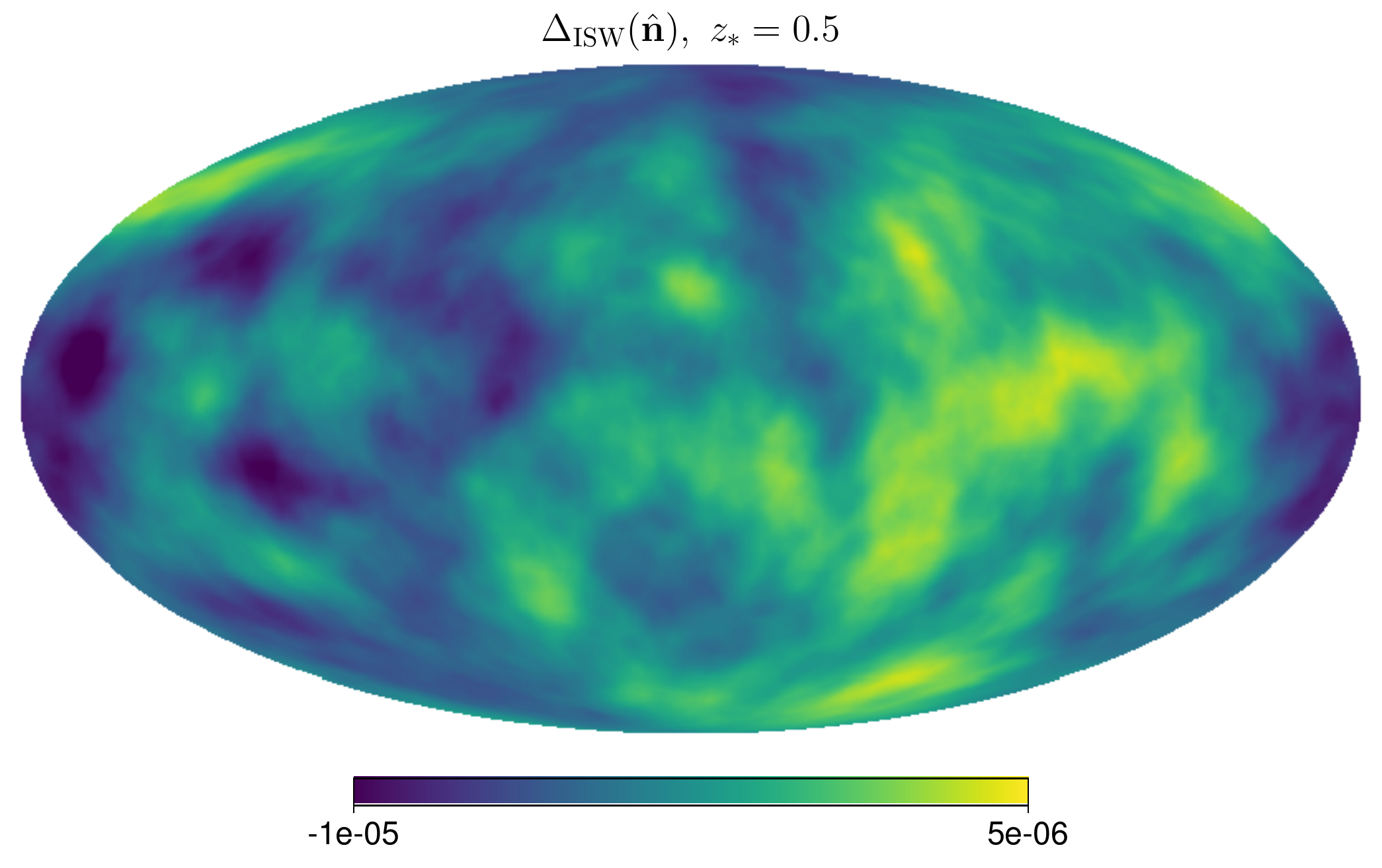}
        \includegraphics[width=0.8\textwidth]{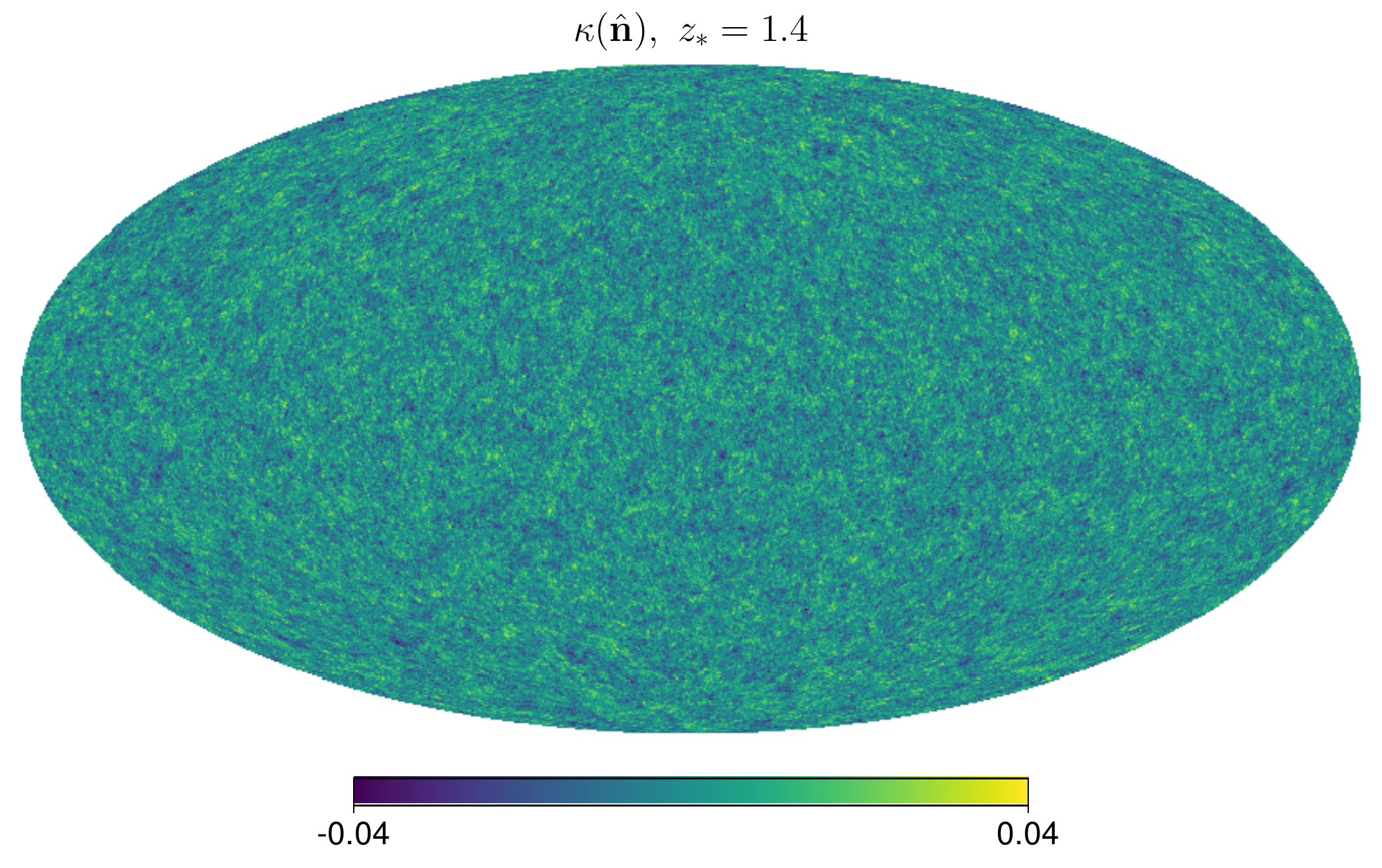}
        \caption{{\sl Top:} simulated map of the low-redshift ISW effect for a source plane at $z_*=0.5$. {\sl Bottom:} map of the lensing convergence for a source plane at $z_*=1.4$.}\label{fig:kisw}
      \end{figure}
      The fluctuations in the gravitational potential perturb the trajectories of photons via gravitational lensing. The effects of gravitational lensing on different cosmological observables are encoded in the so-called ``lensing potential'', defined as \citep{2001PhR...340..291B,2006PhR...429....1L}
      \begin{equation}
        \psi(\nv)\equiv-2\int_0^{\chi_*}d\chi\frac{\chi_*-\chi}{\chi_*\chi}\phi_N(z(\chi),\chi\nv).
      \end{equation}
      The trajectories of photons are deflected by an angle $\boldsymbol{\alpha}\equiv\nabla_{\nv}\psi$, and the shapes of background objects are distorted via the lensing distortion tensor $\mathsf{\Gamma}\equiv-{\sf H}_{\nv}\psi$, where $\nabla_{\nv}$ and ${\sf H}_{\nv}$ are the gradient and Hessian operators on the sphere. The distortion tensor $\mathsf{\Gamma}$ is commonly decomposed into its spin-0 and spin-2 components, the {\sl convergence} $\kappa$ and {\sl shear} $(\gamma_1,\gamma_2)$:
      \begin{equation}
        \mathsf{\Gamma}\equiv\left(
        \begin{array}{cc}
          \kappa+\gamma_1 & \gamma_2 \\
          -\gamma_2 & \kappa-\gamma_1
        \end{array}
        \right).
      \end{equation}

      Computing $\boldsymbol{\alpha}$ and $\mathsf{\Gamma}$ from $\psi$ would require first creating a map of $\psi$ to then differentiate. This is not feasible when calculating the effects of lensing on a large number of sources at different redshifts (see Section \ref{sssec:meth.tr.src}). Instead, we rewrite these quantities as
      \begin{equation}\label{eq:lens}
        \boldsymbol{\alpha}(\nv)=-2\int_0^{\chi_*}d\chi\frac{\chi_*-\chi}{\chi_*}\nabla_\perp \phi_N(z,\chi\nv),
        \hspace{12pt}        \mathsf{\Gamma}(\nv)=-2\int_0^{\chi_*}d\chi\chi\frac{\chi_*-\chi}{\chi_*}{\sf H}_\perp \phi_N(z,\chi\nv),
      \end{equation}
      where $\nabla_\perp$ and ${\sf H}_\perp$ are the gradient and Hessian operators projected onto the plane perpendicular to $\nv$.

      Explicitly, if $\nv\equiv(\sin\theta\,\cos\varphi,\sin\theta\,\sin\varphi,\cos\theta)$, defining the projector
      \begin{equation}
        {\sf P}_{\nv}\equiv\left(
        \begin{array}{ccc}
          \cos\theta\,\cos\varphi & \cos\theta\,\sin\varphi & -\sin\theta \\
          -\sin\varphi & \cos\varphi & 0
        \end{array}
        \right),
      \end{equation}
      the projected gradient and Hessian are
      \begin{equation}
        \nabla_\perp\phi_N\equiv{\sf P}_{\nv}\nabla\phi_N,\hspace{12pt} {\sf H}_\perp\phi={\sf P}_{\nv}({\sf H}\phi_N){\sf P}_{\nv}^T,
      \end{equation}
      where $\nabla_i\equiv\partial/\partial x^i$ and ${\sf H}_{ij}\equiv \partial^2/\partial x^i\partial x^j$.

      The calculation of lensing-related quantities in \colore{} is thus analogous to the procedure outlined in Section \ref{sssec:meth.tr.proj}: along a given line of sight $\nv$ (corresponding to a map pixel or to the position of a given source), the values of the first or second-order derivatives of $\phi_N$ are calculated and interpolated from the Cartesian grid onto a set of equidistant points along $\nv$. The corresponding quantities are then projected onto the plane perpendicular to $\nv$, and the integrals in Eq. \ref{eq:lens} are computed as direct sums over the evaluated points.

      Besides providing lensing information associated with its source catalogs (see Section \ref{sssec:meth.tr.src}), \colore{} returns maps of the lensing convergence $\kappa(\nv)$ for an arbitrary number of source planes at different redshifts. An example at $z_*=1.4$ is shown in the bottom panel of Fig. \ref{fig:kisw}.
    
    \subsubsection{Sources}\label{sssec:meth.tr.src}
      \begin{figure}
        \centering
        \includegraphics[width=0.49\textwidth]{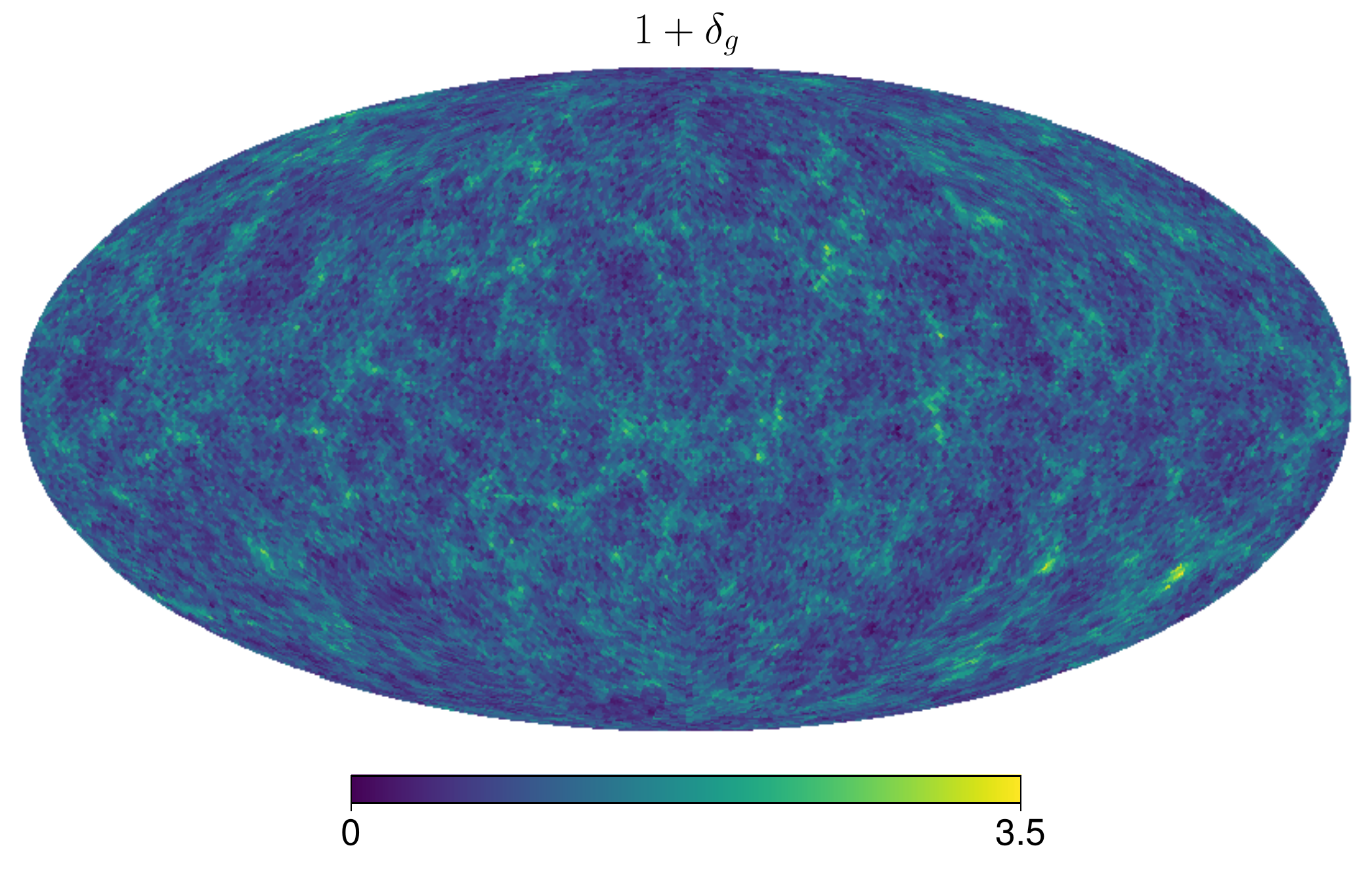}
        \includegraphics[width=0.49\textwidth]{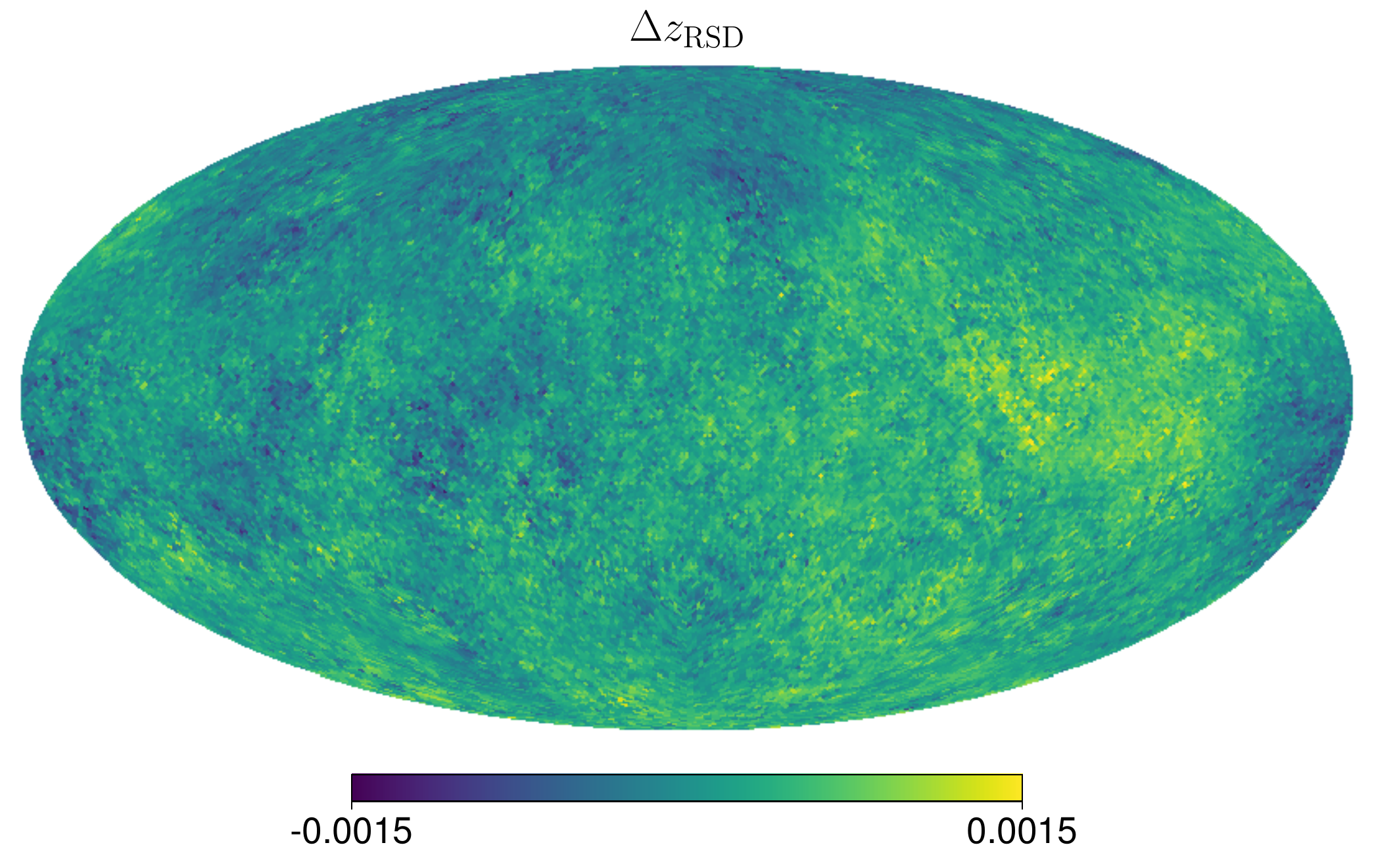}
        \includegraphics[width=0.49\textwidth]{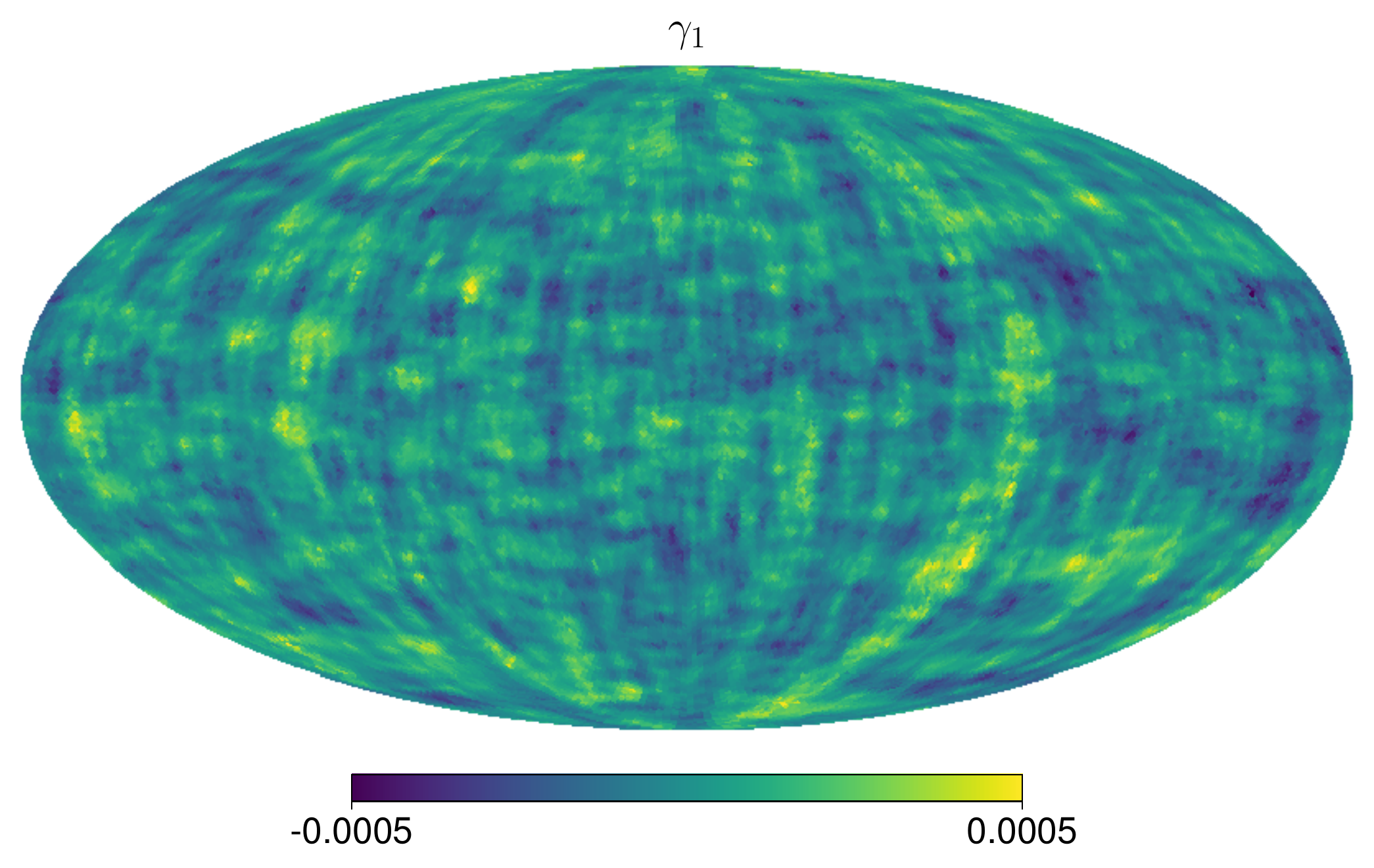}
        \includegraphics[width=0.49\textwidth]{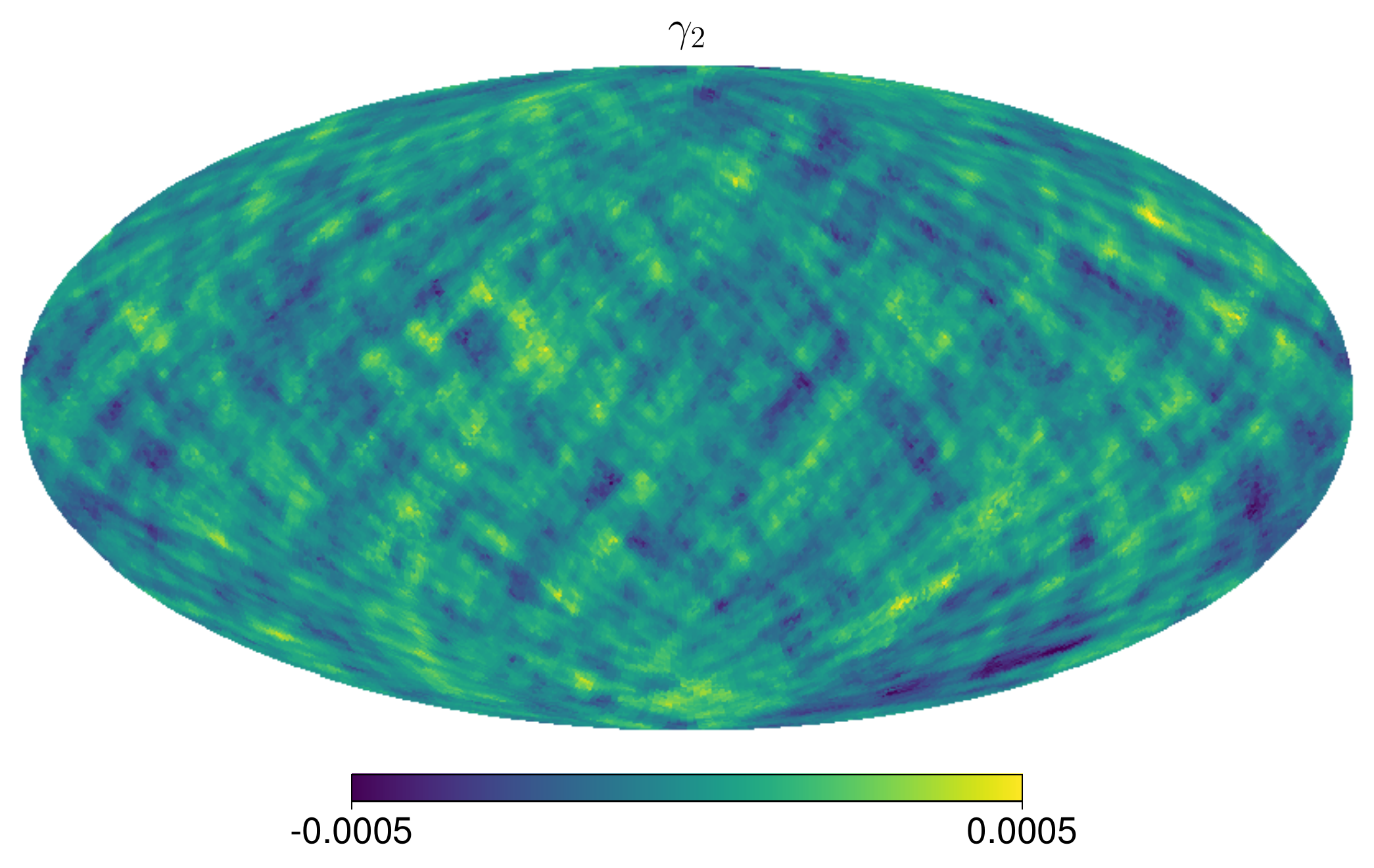}
        \includegraphics[width=0.49\textwidth]{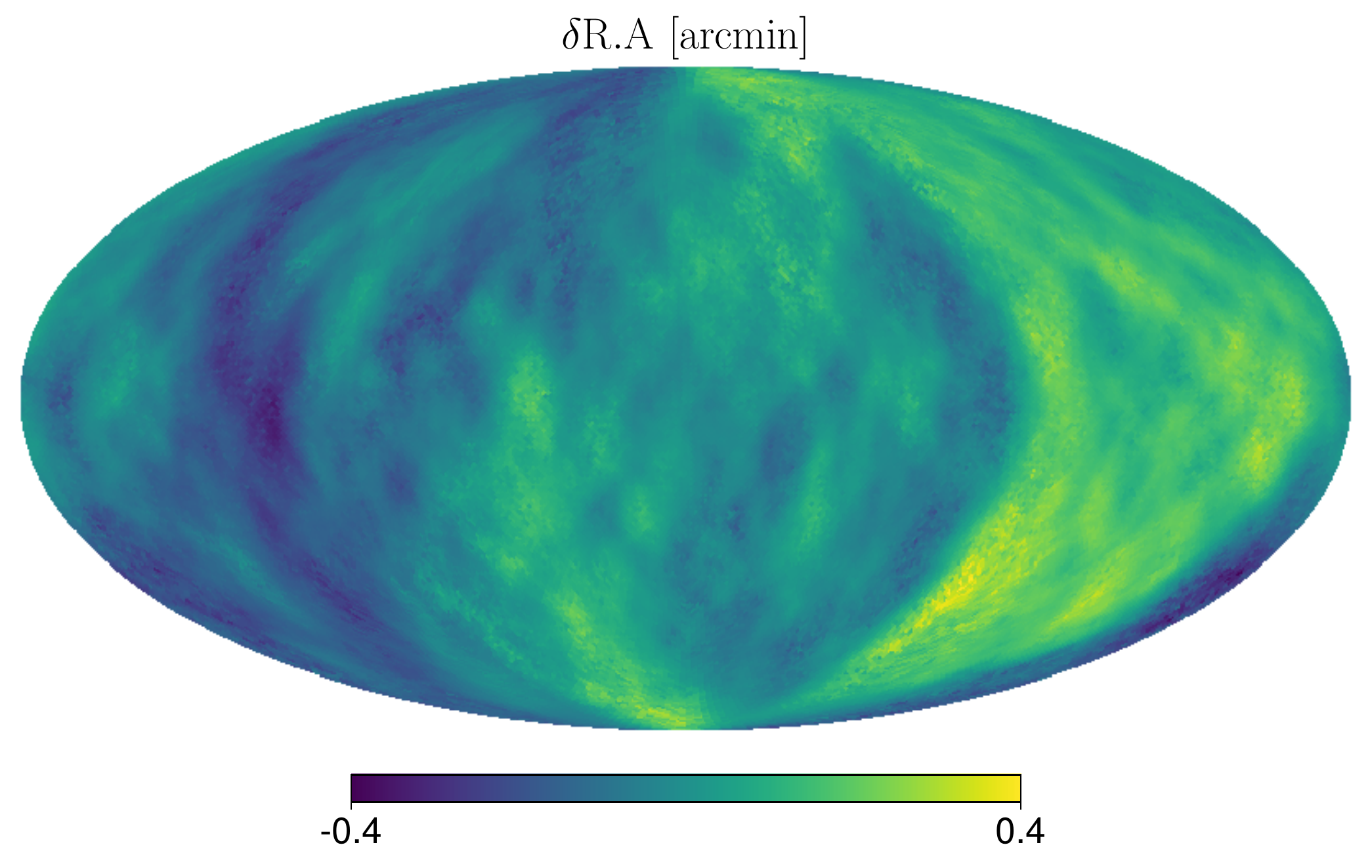}
        \includegraphics[width=0.49\textwidth]{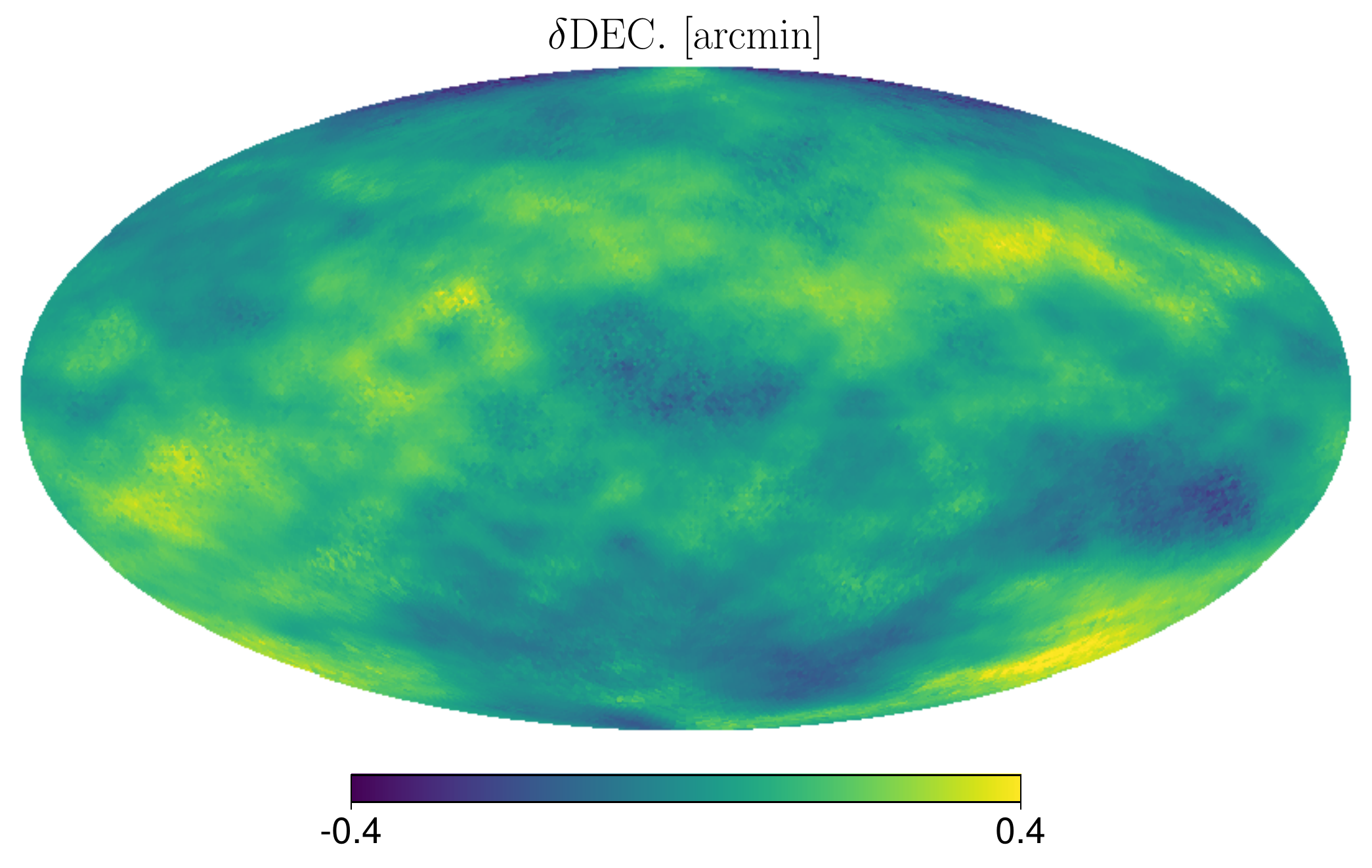}
        \caption{Maps of source-related quantities for a simulated \colore{} catalog in the redshift range $z<0.3$. {\sl Top left:} source overdensity. {\sl Top right:} mean redshift distortion. {\sl Middle:} mean lensing shear. {\sl Bottom:} mean lensing displacement vector.}\label{fig:srcs}
      \end{figure}

      \begin{figure}
        \centering
        \includegraphics[width=0.8\textwidth]{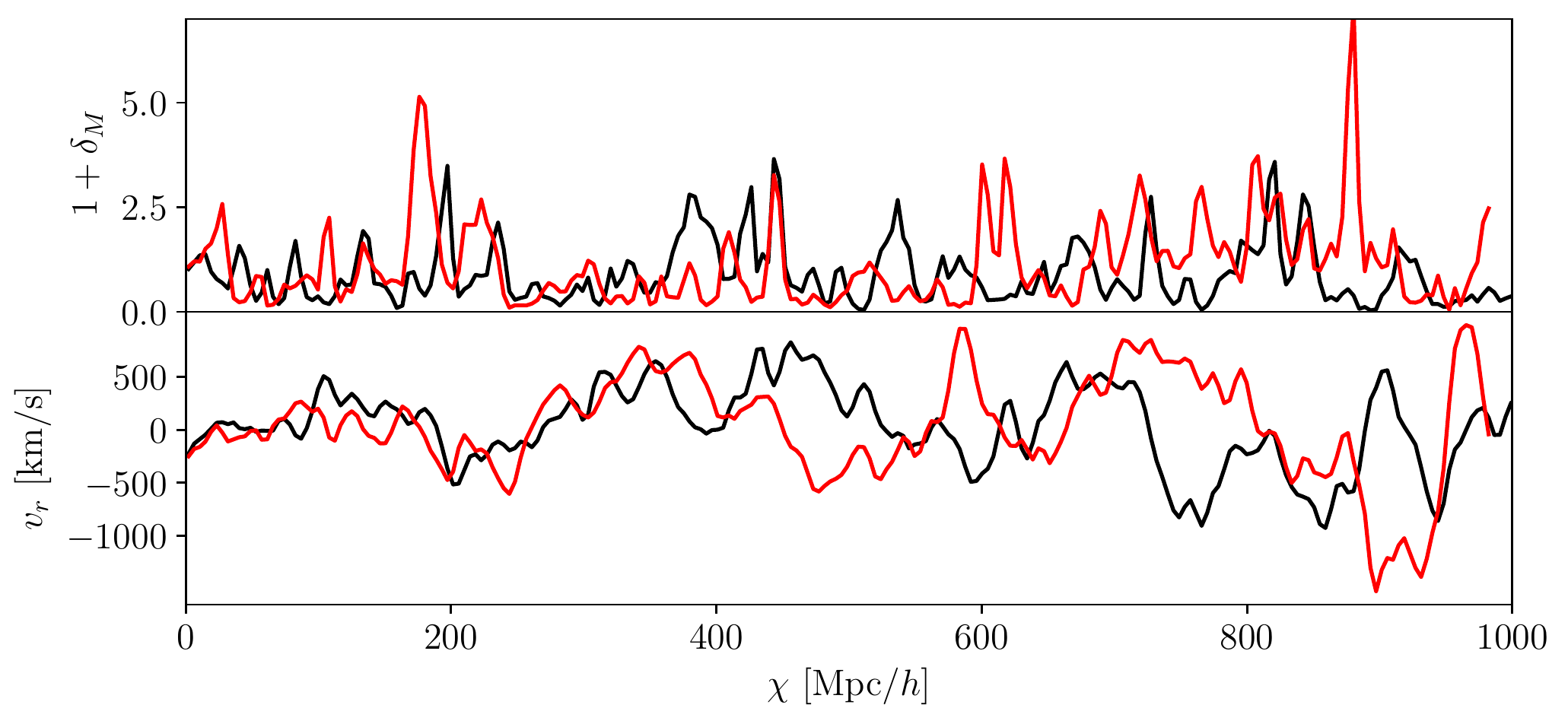}
        \caption{Density ({\sl top}) and radial velocity ({\sl bottom}) skewers for two arbitrary sources in a simulated \colore{} catalog.}\label{fig:skwr}
      \end{figure}

      \colore{} can also generate catalogs of discrete sources as biased tracers of the matter distribution. The source distribution is modelled as a Cox process: the number of sources of type $a$ in a given Cartesian cell $i$, $N_i^a$ is a random Poisson variable with a stochastic mean given by 
      \begin{equation}
        \bar{N}_i^a = (\Delta x)^3\bar{n}_a(\chi)\frac{B_a(\delta_M({\bf x}_i))}{\langle B_a(\delta_M)\rangle},
      \end{equation}
      where ${\bf x}_i$ are the coordinates of cell $i$, $(\Delta x)^3$ is the cell volume, $\bar{n}_a(\chi)$ is the redshift-dependent mean density of sources, and $B_a$ is the biasing relation of type-$a$ sources (see Eqs. \ref{eq:bias_1}-\ref{eq:bias_3}).

      Once $N_i^a$ has been determined for each cell, the corresponding number of sources are generated with three-dimensional positions randomly located within their cell. The three-dimensional comoving position is then translated into observable angular and redshift coordinates. In addition to the cosmological redshift, the contribution from the source's peculiar velocity, necessary to simulate redshift-space distortions, is calculated as $\Delta z_{\rm RSD}=v_r$, where the radial velocity is calculated from the gradient of the gravitational potential as
      \begin{equation}
        v_r(z,{\bf x})=-\frac{2}{3H_0^2\Omega_M}f(z)\,(\nv\cdot\nabla)\phi_N(z,{\bf x}).
        \label{eq:radial_velocity}
      \end{equation}
	  Given this derivation of velocities they will only include linear effects.

      If desired, the source catalogs can also contain gravitational lensing information ($\alpha_\theta$, $\alpha_\varphi$, $\kappa$, $\gamma_1$, $\gamma_2$) for each source. This can be used to construct a weak lensing shear catalog with ellipticities $e_i=\gamma_i$, or to include the effects of lensing magnifications by perturbing the source angular positions $\theta\rightarrow\theta+\alpha_\theta$, $\varphi\rightarrow\varphi+\alpha_\varphi$, and its flux by $F\rightarrow F(1+2\kappa)$. In this case, all lensing quantities are calculated as described in Section \ref{sssec:meth.tr.gl}, by integrating the interpolated transverse derivatives of the gravitational potential along each source's line of sight.

      Source catalogs can also be endowed with so-called ``line-of-sight skewers'', containing the matter overdensity and radial velocity interpolated from the Cartesian box onto each source's line of sight. As in the case of gravitational lensing calculations, trilinear interpolation is used, and both fields are sampled at radial comoving intervals equal to \colore{}'s Cartesian cell size. The use of skewers to produce simulated observations of the Lyman-$\alpha$ forest was discussed in detail in \cite{1912.02763}.

      It is worth noting that the computation of any quantity requiring full line-of-sight information (lensing or skewers) requires redistributing sources from ``slabs'' into ``beams'' across nodes, and for all nodes to loop through the full Cartesian box to add the contribution of all slabs to their beams (i.e. points 7 and 8 in Section \ref{ssec:meth.code}). This process requires significant communication between MPI nodes and, depending on the particular case, can have a significant impact on the total computing time. For this reason, if no line-of-sight information is requested from \colore{} (i.e. if only sources or intensity maps are requested, with no associated lensing information or skewers), steps 7 and 8 are skipped, often leading to a significant speed-up.

      Furthermore, computing the lensing observables along each line-of-sight for large catalogs containing billions of sources, as would be the case e.g. if simulating the full LSST shear sample, is a computationally demanding task that can dominate by far all other steps in a \colore{} run. To avoid this, an alternative approximate scheme to compute these quantities can be used. In the ``fast lensing'' scheme, the five lensing observables (displacement, convergence and shear) are precomputed on a set of spherical {\tt HEALPix} maps at constant radial comoving distance intervals, using the same method outlined in Section \ref{sssec:meth.tr.gl}. The angular resolution parameter $N_{\rm side}$ of each map is adaptively chosen so that the physical pixel size is smaller than the Cartesian cell size, in order to avoid oversampling as well as degrading the original three-dimensional resolution. The hierarchical nature of the {\tt HEALPix} scheme makes it possible to relate a pixel in a given map with pixels in all lower-resolution maps with smaller radii. The lensing observables for a given source are then calculated by interpolating between the values of that observable along the line of pixels corresponding to the source's angular coordinates.

      Figure \ref{fig:srcs} shows maps of the quantities described in this section for a small source catalog simulated with \colore{}. The simulated sample had a redshift distribution
      \begin{equation}
        \frac{dN}{dz}\propto\left(\frac{z}{z_0}\right)^2\exp\left[-\left(\frac{z}{z_0}\right)^{3/2}\right],
      \end{equation}
      with $z_0=0.07$, extending up to $z\lesssim0.3$. The simulation was run using the first-order LPT structure formation model. The different panels show the source overdensity and mean redshift distortion (top panels), the mean lensing shear (middle panels) and the mean lensing deflection vector (bottom panels). Figure \ref{fig:skwr} shows density and velocity skewers (top and bottom panels respectively) for two arbitrary sources in the same catalog.

    \subsubsection{Line intensity mapping}\label{sssec:meth.tr.imap}
      \begin{figure}
        \centering
        \includegraphics[width=0.99\textwidth]{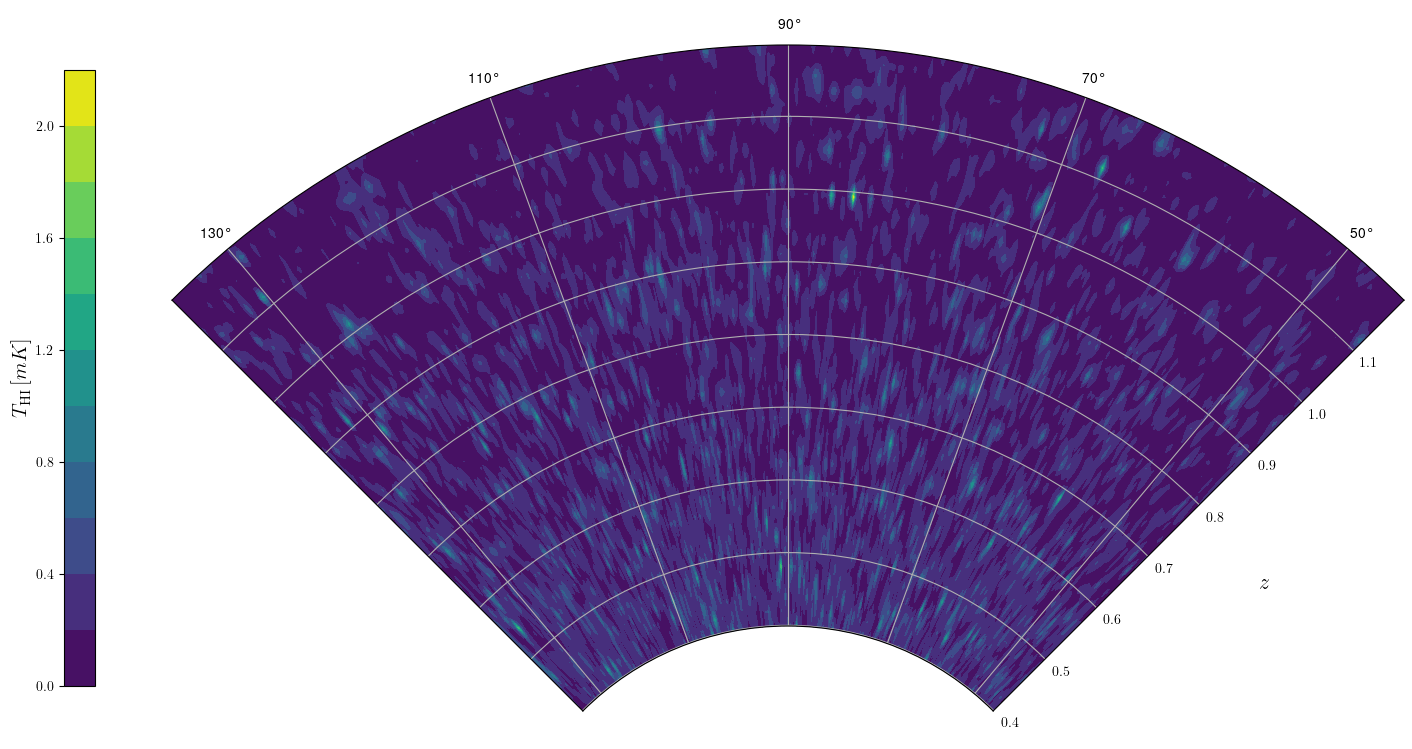}
        \caption{Slice through one of the 21cm intensity maps generated with \colore{}.}\label{fig:imap}
      \end{figure}
      Consider a species of gas emitting at a rest-frame frequency $\nu_0$ due to some atomic or molecular transition. The intensity (flux per unit frequency) measured in a patch around $\nv$ with solid angle $\delta\Omega$, and in a frequency interval $\delta\nu$ around $\nu$ is given by \cite{2005MNRAS.360...27A}
      \begin{equation}
        I(\nu,\nv)=\frac{\hbar A_{21} \nu_0 x_2}{2m_a}\frac{M_{\rm em}}{(1+z)^2\chi^2\delta\Omega\,\delta\nu},
      \end{equation}
      where $A_{21}$ is the Einstein coefficient for the transition, $m_a$ is the atomic mass of the emitting gas, $x_2$ is the fraction of the gas in the excited state, $M_{\rm em}$ is the total mass of the emitting gas in the voxel defined by $(\delta\nu,\delta\Omega)$. Associating this intensity with a black-body temperature in the Rayleigh-Jeans regime ($T=c^2 I/(2k_B\nu^2)$), this can be rewritten as:
      \begin{equation}
        T(\nu,\nv)=\bar{T}(z)\left[1+\delta_{\rm em}(z,\nv)\right], 
      \end{equation}
      where $\delta_{\rm em}$ is the overdensity of the emitting gas in redshift space (i.e. accounting for the effects of redshift-space distortion), $z=\nu_0/\nu-1$, and the mean temperature $\bar{T}$ is
      \begin{equation}
        \bar{T}(z)\equiv\frac{3\,\hbar\,A_{21}\,x_2\,x_{\rm em}(z)\,\Omega_{b,0}\,H_0^2\,c^2\,(1+z)^2}{32\pi\,G\,k_B\,m_a\,\nu_0^2\,H(z)}.
      \end{equation}
      
      \colore{} generates mock intensity mapping observations by modelling $\delta_{\rm em}$ as a biased tracer of $\delta_M$ using one of the bias relations described in Section \ref{sssec:meth.tr.proj}. The code estimates the mean brightness temperature in each Cartesian voxel in terms of the value of the matter overdensity, and a redshift-dependent mean temperature and bias. In order to interpolate from the Cartesian grid on to temperature maps, while accounting for the effects of redshift space distortions, each cell is first sub-divided into smaller sub-cells. The Cartesian coordinates of each sub-cell are translated into angular coordinates $(\theta,\varphi)$ and emission frequency $\nu=\nu_0/(1+z+v_r)$, where $z$ is the redshift corresponding to the sub-cell's radial comoving distance, and $v_r$ is the value of the radial comoving peculiar velocity field. Each sub-cell is then assigned to a given pixel and frequency band based on $(\theta,\varphi,\nu)$. The final intensity maps are then generated by averaging over the temperature of all sub-cells thus assigned to each pixel.

      The implementation used by \colore{} is very similar to that used by {\tt CRIME} \cite{2014MNRAS.444.3183A}. \colore{} does not emulate any other secondary effects, such as second-order gravitational lensing \cite{2015MNRAS.448.2368P,2018PhRvD..97l3539S} or self-absorption. If desired, gravitational lensing could be simulated by perturbing the angular positions of the Cartesian sub-cells above with the deflection field described in Section \ref{sssec:meth.tr.gl}. Figure \ref{fig:imap} shows a slice through a set of intensity maps corresponding to the 21cm line simulated using the model described in \cite{2014MNRAS.444.3183A} in the range $\nu\in[1015, 646]\,\,{\rm MHz}$ ($0.4<z<1.2$).

  \subsection{Lognormal predictions}\label{ssec:meth.lognormal}
    Although, as discussed in Section \ref{sssec:meth.box.ln}, the log-normal transformation is not able to recover the right properties of the non-linear density fluctuations at all orders, one of its key advantages is the possibility to produce exact analytical predictions for the two-point correlators of a lognormal field. This makes it possible to compare the output of a \colore{} simulation against precise predictions, making it straightforward to, for example, quantify the impact of different observational systematic effects added in post-processing on the main observables of a large-scale structure experiment.

    The real-space two-point correlation function of a lognormal field is related to that of its parent Gaussian field via:
    \begin{equation}\label{eq:xiln}
      1+\xi_{\rm LN}(r)=\exp\left[\xi_{\rm G}(r)\right],
    \end{equation}
    where $\xi_{\rm LN}(r)\equiv\langle\delta_{\rm LN}({\bf x})\delta_{\rm LN}({\bf x}+{\bf r})\rangle$ and $\xi_{\rm G}(r)\equiv\langle\delta_{\rm G}({\bf x})\delta_{\rm G}({\bf x}+{\bf r})\rangle$. Its cross-correlation with the Gaussian field itself is simply
    \begin{equation}
      \langle\delta_{\rm G}({\bf x})\delta_{\rm LN}({\bf x}+{\bf r})\rangle=\xi_{\rm G}(r).
      \label{eq:xc_gaussian_lognormal}
    \end{equation}
    Computing the power spectrum of the lognormal field $P_{\rm LN}(k)$ from that of the parent Gaussian field $P_{\rm G}(k)$ is thus a simple three-step process:
    \begin{enumerate}
      \item Compute the Gaussian correlation function from $P_{\rm G}(k)$ via
      \begin{equation}
        \xi_{\rm G}(r)=\frac{1}{2\pi^2}\int_0^\infty dk\,k^2\,P_{\rm G}(k)\,\frac{\sin(kr)}{kr}.
      \end{equation}
      \item Compute $\xi_{\rm LN}$ from $\xi_{\rm G}$ through Eq. \ref{eq:xiln}.
      \item Compute the $P_{\rm LN}(k)$ from $\xi_{\rm LN}(r)$ via
      \begin{equation}
        P_{\rm LN}(k)=4\pi\int_0^\infty dr\,r^2\,\xi_{\rm LN}(r)\,\frac{\sin(kr)}{kr}.
      \end{equation}
    \end{enumerate}

    \colore{} automatically produces and outputs predictions for the power spectrum and correlation function of any lognormal biased tracers simulated. These predictions are valid as long as the simulation was run using the lognormal structure formation model, and the exponential bias model (see Section \ref{sssec:meth.tr.proj}), which retains the lognormal nature of the biased fields.

    It is worth noting that, in order to produce truly accurate predictions, it is necessary to account for all sources of smoothing produced by the different operations carried out by \colore{}:
    \begin{itemize}
      \item The chosen Gaussian smoothing scale.
      \item The finite resolution of the Cartesian grid.
      \item The effects of interpolating from the Cartesian grid onto beam-related quantities (line-of-sight integrals, skewers etc.).
    \end{itemize}
    The latter two effects (finite grid resolution and interpolation), have an associated Fourier-space window function given by
    \begin{equation}
      W_n({\bf k})=\left[\left(2\frac{\sin(2k_x\Delta x)}{k_x\Delta x}\right)\left(2\frac{\sin(2k_y\Delta x)}{k_y\Delta x}\right)\left(2\frac{\sin(2k_z\Delta x)}{k_z\Delta x}\right)\right]^n,
    \end{equation}
    where $\Delta x$ is the grid spacing, and $n=1$ and 2 for nearest-neighbour and trilinear interpolation, respectively. For $k\ll1/\Delta x$, we can approximate $W_n$ as a Gaussian filter via
    \begin{equation}
      W_n({\bf k})\simeq\left[1-\frac{(k_x^2+k_y^2+k_z^2)(\Delta x)^2}{24}\right]^n\simeq e^{-(k R_G)^2/2}.
    \end{equation}
    where $R_G^2\equiv n(\Delta x)^2/12$. Thus, the effective smoothing scale associated to the appropriate number of interpolation operations can be simply added in quadrature to the chosen Gaussian smoothing scale to produce theoretical predictions.

\section{Results}\label{sec:res}
  \subsection{Validation}\label{ssec:res.val}
    \begin{figure}
      \centering
      \includegraphics[width=0.65\textwidth]{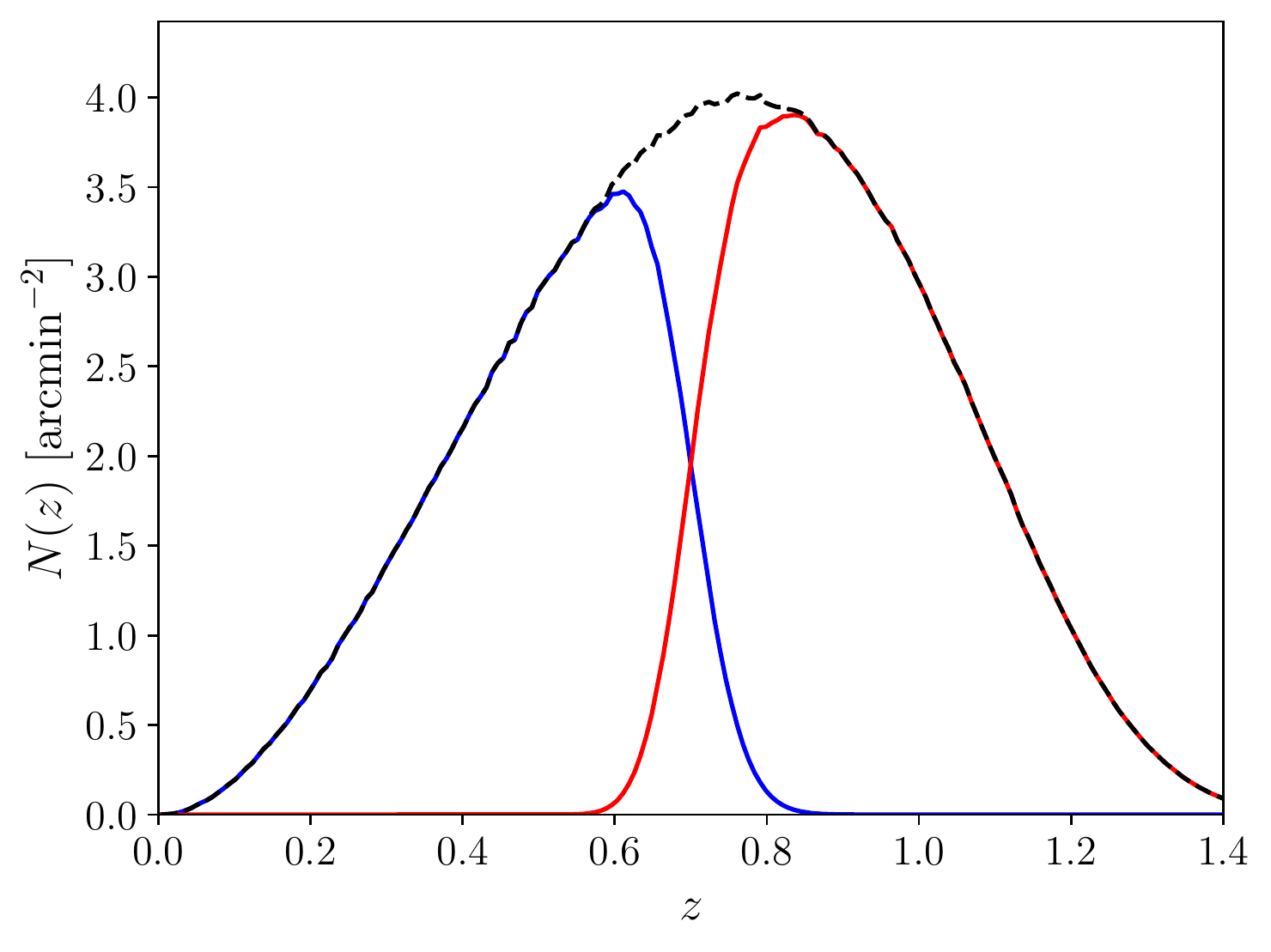}
      \caption{Redshift distribution of the two tomographic bins used in the analysis of the validation simulations (red and blue lines), as well as the overall redshift distribution (black dashed line). The bins are defined by a cut in photometric redshift space at $z_{\rm photo}=0.7$, where we assigned each source a random photometric redshift error with standard deviation $\sigma_z=0.03\,(1+z)$.} \label{fig:dndz_val}
    \end{figure}
    \begin{figure}
      \centering
      \includegraphics[width=0.99\textwidth]{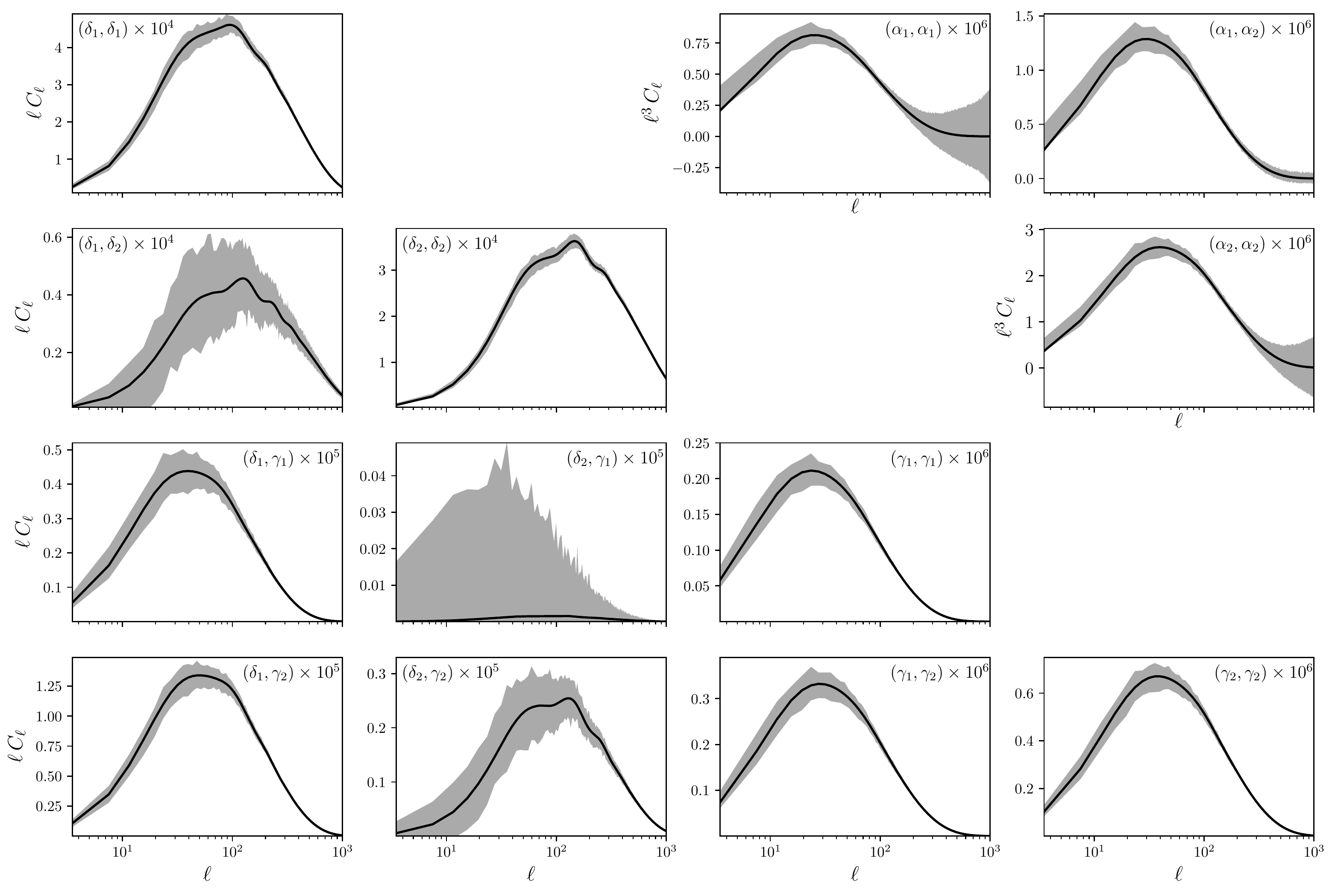}
      \caption{Galaxy clustering, cosmic shear and lensing displacement power spectra. The gray bands show the 68\% scatter from the 100 validation realizations, and the theoretical predictions, described in Section \ref{ssec:meth.lognormal}, are shown as black solid lines. The lower-left corner shows all auto- and cross-correlations between the two galaxy clustering and cosmic shear bins. The upper-right corner shows the auto- and cross-correlations between the lensing displacement vectors in both redshift bins.} \label{fig:cl_txt}
    \end{figure}
    \begin{figure}
      \centering
      \includegraphics[width=0.99\textwidth]{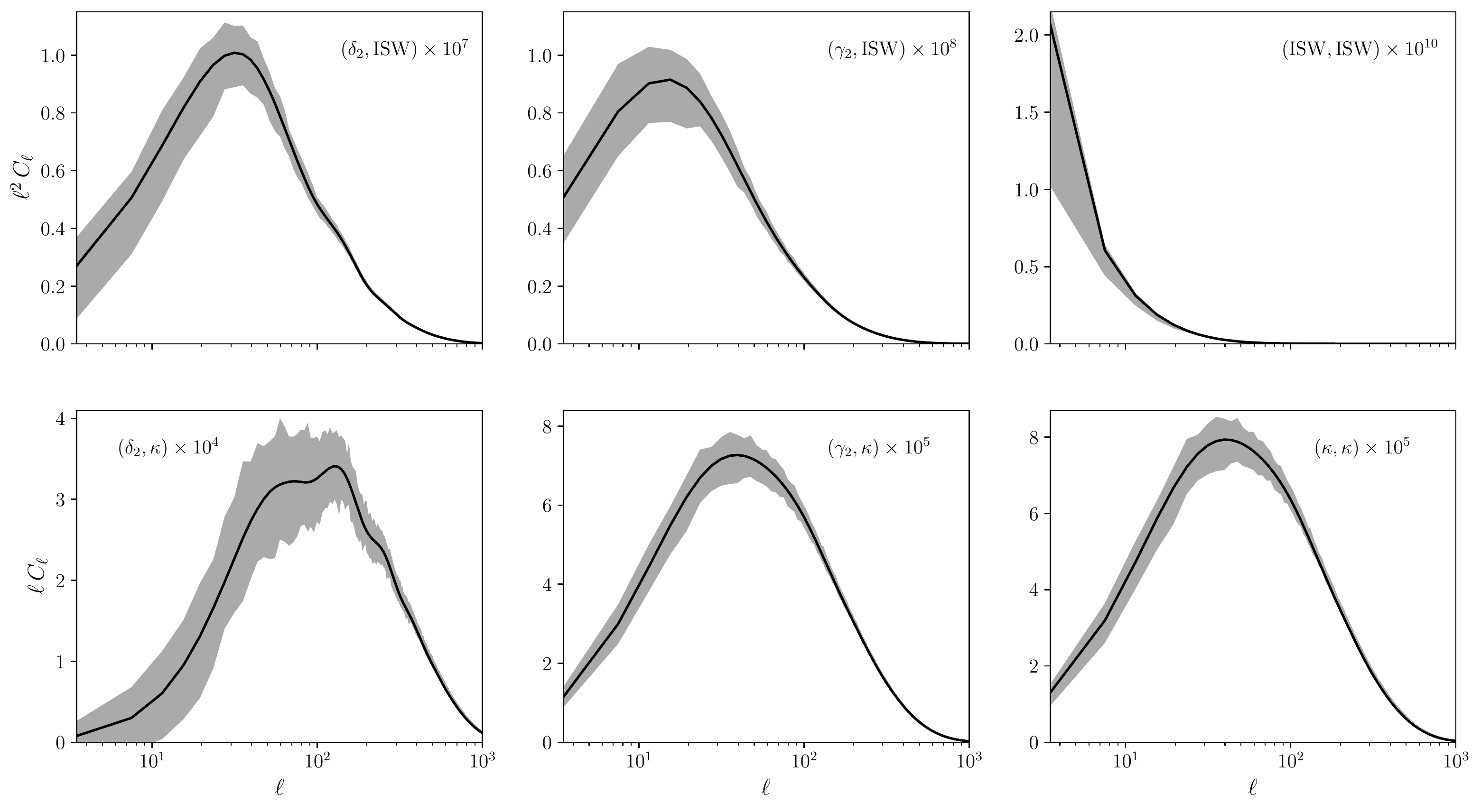}
      \caption{Cross-correlations between ISW (top row) and convergence maps (bottom row) at $z=1$ and the two high-redshift clustering and shear samples in the validation simulations. The gray bands show the 68\% scatter from the 100 validation realizations, and the theoretical predictions, described in Section \ref{ssec:meth.lognormal}, are shown as black solid lines.} \label{fig:cl_txt_ik}
    \end{figure}
    \begin{figure}
      \centering
      \includegraphics[width=0.7\textwidth]{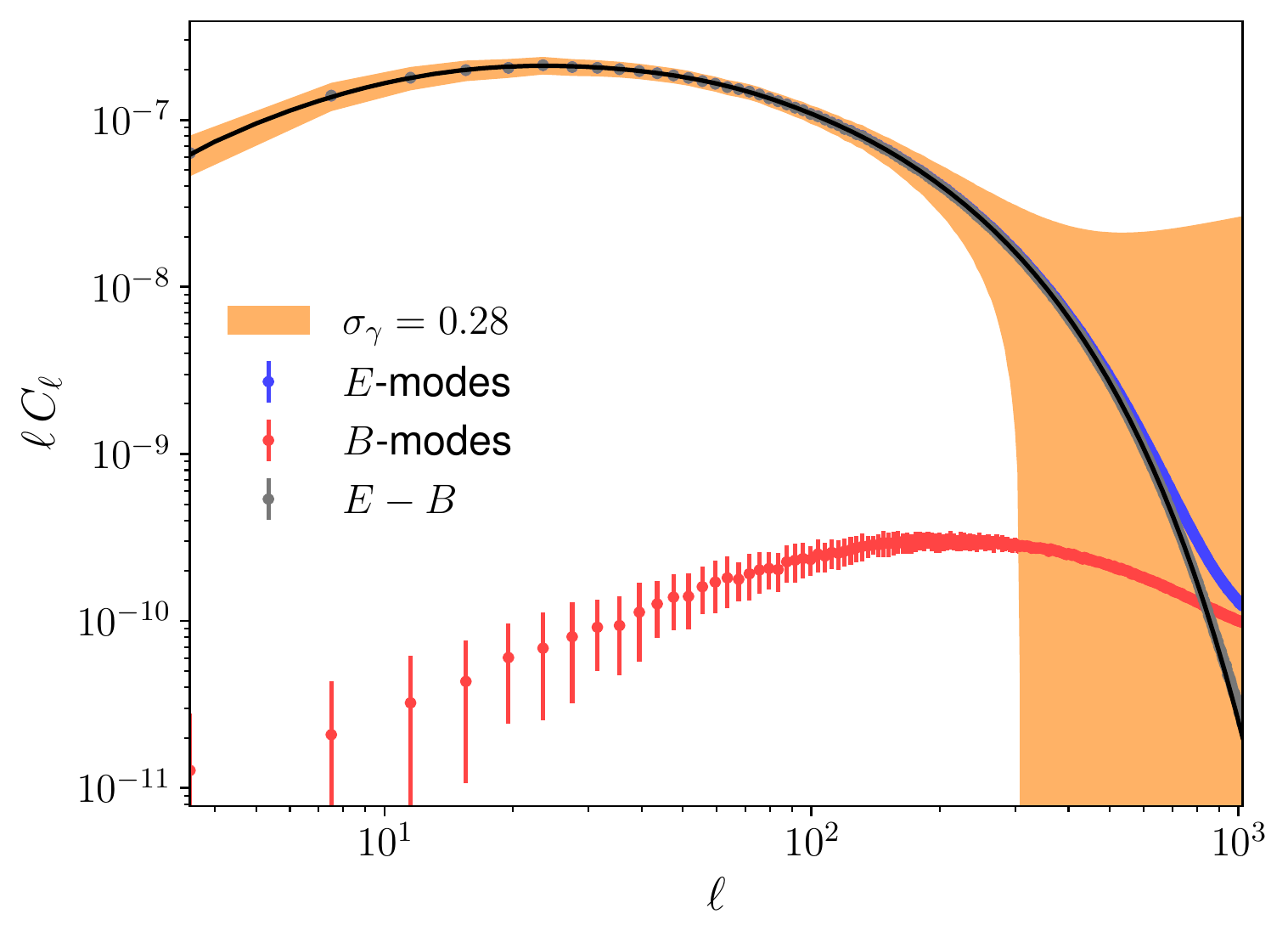}
      \caption{Shear power spectra for the first redshift bin in the validation simulations. The blue and red points show the results (mean and standard deviation of the 100 simulations) with no correction for the effects of source clustering. The gray points show the result of applying this correction by simply subtracting the $B$-mode power spectrum from the $E$-mode one. The black line shows the theoretical prediction for the latter. The orange band shows the 1$\sigma$ uncertainties one would find in the presence of realistic shape noise, which would make the source clustering effect undetectable in practice.} \label{fig:clbb}
    \end{figure}
    \colore{} has been the basis of other analyses, and some of its functionality was validated in previous work. The ability to produce large-scale intensity maps for the 21cm neutral hydrogen line was presented, used, and validated in \citep{2014MNRAS.444.3183A,1609.00019}, and \citep{1808.03093} used the code to explore cross-correlations with galaxy clustering data. 
    \cite{1912.02763} used \colore's line-of-sight skewers to produce mock observations of the Lyman-$\alpha$ forest; these mocks were extensively used to validate the final Lyman-$\alpha$ BAO analysis of the eBOSS collaboration \cite{2020ApJ...901..153D}.
    Our discussion here therefore focuses on presenting and validating \colore{} as a tool to produce fast simulations for wide galaxy surveys targeting galaxy clustering (both spectroscopic and photometric), weak lensing shear, and their cross-correlation with maps of the lensing convergence, and the ISW effect.

    To do so, we have run a set of 100 large \colore{} realisations containing these observables, and compared the relevant two-point correlations from different pairs of tracers in the simulation with the corresponding theoretical predictions.

    \subsubsection{Simulations}\label{sssec:res.val.sim}
      We generate 100 realisations covering the volume up to redshift $z=1.4$, corresponding to a box size $L_{\rm box}=5{,}843\,{\rm Mpc}/h$. We use a grid of size $N_{\rm grid}=2048$, with cell size $\Delta x=2.85\,{\rm Mpc}/h$. Each realisation is populated with a single galaxy sample with the redshift distribution shown in Fig. \ref{fig:dndz_val}, and a total number density $\bar{n}_g=2.9\,{\rm arcmin}^{-2}$. We used a lognormal structure formation model, and an exponential bias model with bias $b(z)=1+0.65z+0.03z^3$, compatible with a blue galaxy sample \citep{2006A&A...448..101G}. We also used \colore{} to generate maps of the lensing convergence and the ISW effect at $z=1$. We used cosmological parameters $\Omega_m=0.3$, $\Omega_b=0.05$, $h=0.7$, $n_s=0.96$, $\sigma_8=0.8$. The Gaussian overdensity field was smoothed with a Gaussian kernel with width $R_G=2\,{\rm Mpc}/h$.

      In order to simulate the effects of redshift uncertainties in photometric redshift surveys, we assigned a random Gaussian error to each galaxy redshift with standard deviation $\sigma_z=0.03\,(1+z)$. We then divided all galaxies into two redshift bins, corresponding to sources above and below redshift $z_{\rm thr}=0.7$. The redshift distributions of the resulting bins are shown in Fig. \ref{fig:dndz_val}.
      For each redshift bin, we created maps of the projected galaxy overdensity, as well as the shear and lensing displacement vectors. We did not make use of the fast lensing scheme described in Section \ref{sssec:meth.tr.src} for these simulations. The latter quantities are provided by \colore{} at each source. Finally, we computed all auto- and cross-power spectra between these maps, as well as the corresponding correlations with the lensing convergence and ISW maps at $z=1$ produced by \colore{}. In order to optimally weight the shear and displacement fields by the local number of sources when computing these power spectra (and in order to account for their spin-2 and spin-1 nature), we make use of {\tt NaMaster} \citep{2019MNRAS.484.4127A}.
    
      For the 3D clustering validation we used 10 out of the 100 realisations. Since the aim in this case is simulating a spectroscopic survey, we do not add photometric redshift uncertainties. We used the \texttt{corrfunc}~\citep{corrfunc} package to obtain measurements of the monopole and the quadrupole correlation functions for two different redshfit bins $(0.5, 0.7)$ and $(0.7, 0.9)$; probing separations between $r=0.1 \ {\rm Mpc}/h$ and $r=200 \ {\rm Mpc}/h$ in 41 linearly spaced bins.

    \subsubsection{Validation of 2D observables}\label{sssec:res.val.2d}
    
      Before presenting the results from this validation exercise, it is worth clarifying a technical point about the theoretical predictions used to compare with the two-point functions estimated from the simulation. As described in Section \ref{ssec:meth.lognormal}, besides the effects of the lognormal transformation, one must account for the additional smoothing associated with the finite grid and the different interpolation operations. We do so by adding an extra smoothing scale in quadrature to the Gaussian smoothing scale used in the simulation, with the form
      \begin{equation}
        \Delta R_G^2 = n_{\rm eff}\,\frac{(\Delta x)^2}{12},
      \end{equation}
      where the prefactor $n_{\rm eff}$ depends on the finite-resolution effects that must be taken into account. For instance, a single nearest-neighbour interpolation would correspond to $n_{\rm eff}\simeq 1$, while linear interpolation would have $n_{\rm eff}\simeq2$. Thus, since galaxies are assigned to their nearest grid cell, and since two linear interpolations are needed to assign lensing properties to sources, the galaxy clustering and cosmic shear tracers produced by \colore{} should take additional smoothing with $n_{\rm eff}\simeq1$ and $4$ respectively (not taking into account the additional interpolations involved in the fast lensing scheme). Since this additional Gaussian smoothing is not exact, we must adapt these prefactors slightly, in order to improve the agreement between theory predictions and simulations on small scales (high $\ell$s). For instance, we find that the galaxy-galaxy, galaxy-matter, and matter-matter power spectra must be smoothed by additional factors $n_{\rm eff}^{gg}=0.9$, $n_{\rm eff}^{gm}=4$, and $n_{\rm eff}^{mm}=3.8$ to recover the galaxy clustering and cosmic shear power spectra from the \colore{} simulations. Therefore, care must be exercised when interpreting the results of \colore{} simulations, particularly on physical scales smaller than, or comparable with, the grid resolution. The final users are encouraged to tune these $n_{\rm eff}$ factors to their particular analysis, depending on their use case and required level of accuracy.
    
      The gray bands in Figure \ref{fig:cl_txt} show the 1$\sigma$ scatter of the power spectra estimated from the 100 validation simulations, together with the corresponding theoretical prediction in solid black. All auto- and cross-correlations between the projected galaxy overdensity and the cosmic shear field in the two redshift bins described above are shown in the lower-left panels in the figure. The upper-right panels show the three auto- and cross-correlations between the $E$-mode fields corresponding to the lensing displacement vectors in both redshift bins. Similar results are shown in Figure \ref{fig:cl_txt_ik} for correlations involving the lensing convergence and ISW maps, and the projected overdensity and cosmic shear fields in the second. A good agreement between theory and simulations, at the $1\sigma$ level, is found in all cases.

      One further complication must be noted. Cosmic shear is measured at the positions of clustered sources, and this clustering is correlated with the cosmic shear signal itself. This induces an additional contribution to the observed cosmic shear statistics which gives rise to shear $B$-modes, in addition to modifying the $E$-mode power spectrum. This effect is well known, and should be unobservable in most cases \citep{2002A&A...389..729S}. However, in the absence of shape noise (i.e. since we have access to the true weak lensing shear at each source), the effect can be measured in the catalogs produced by \colore{}. This is illustrated in Fig. \ref{fig:clbb}, which shows the $E$ and $B$-mode power spectra for the first redshift bin in the validation simulations without any correction for source clustering (blue and red respectively) in comparison with the theory prediction without source clustering (black line) and the expected uncertainties in the presence of realistic shape noise (orange band). We find that, in practice, the effects of source clustering in the $E$-mode power spectrum can be corrected by simply subtracting the $B$-mode power spectrum from the $E$-mode power spectrum (gray points in the figure).

      \subsubsection{Validation of 3D clustering}\label{sssec:res.val.3d}
\begin{figure}[h]
	    \centering
    	\includegraphics[width=0.95\textwidth]{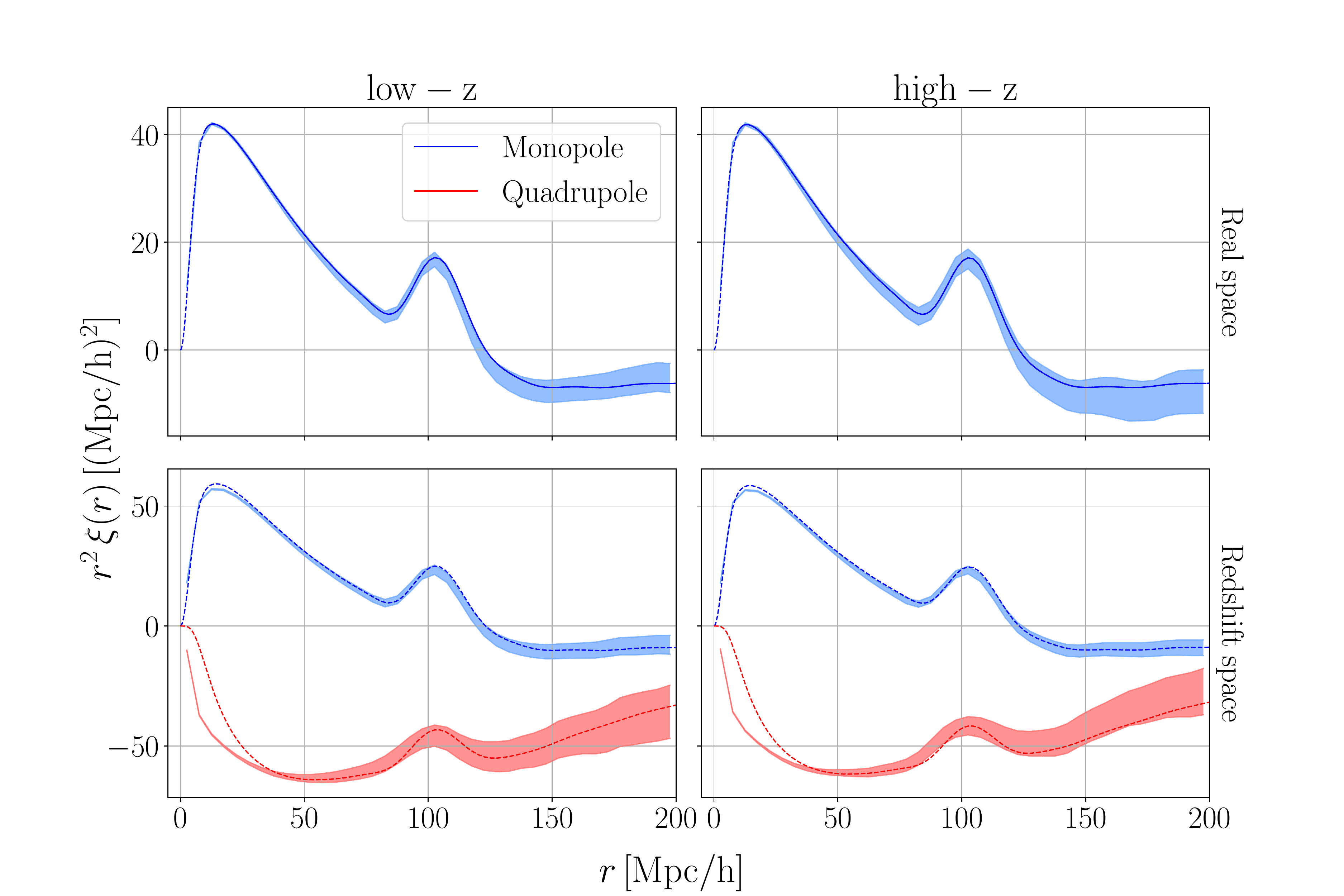}
	    \caption{Measurements of the correlation function from the stack of 10 realizations used to validate the 3D clustering. The low-z samples take redshifts from 0.5 to 0.7, while high-z samples take redshifts from 0.7 to 0.9. The lines show the model, solid in the regions where the bias was fitted, and the shaded bands show the error for a single realization. \emph{Top}: Measurements of the monopole in real space. \emph{Bottom}: Measurements of the monopole and quadrupole in redshift space. 
        }
	    \label{fig:3d_clust}
      \end{figure}

      We validated the 3D clustering of simulated galaxies both in real and in redshift space. 
      The top panel of Figure \ref{fig:3d_clust} shows the real-space monopole for two spectroscopic redshift bins: (0.5-0.7) and (0.7-0.9). The measurement of the correlation function was done separately in 48 different healpix pixels defining the \colore{} beams, and the error bands (estimated from the scatter of these measurements) show the uncertainty for a single realization. Solid lines show the prediction derived in Section \ref{ssec:meth.lognormal} with the linear bias left as a free parameter. The best-fit value of bias (using separations larger than 10 Mpc/h) agrees with the input value at the 2\% level. 
      
      The bottom panel in Figure \ref{fig:3d_clust} shows the redshift-space monopole and quadrupole for the same redshift bins and linear bias.
      We do not have a prediction for the clustering of galaxies in redshift space that is valid on all scales, but we use a modified version of the lognormal model that includes linear redshift-space distortions (Kaiser model \cite{10.1093/mnras/227.1.1}):
      \begin{equation}
    	\label{eq:densityinredshift}
    	\delta_{\rm LN}^s({\bf k}) = \delta_{\rm LN}({\bf k}) + f \mu^2 \delta_G ({\bf k})
      \end{equation}
      where the Gaussian term stands due to the fact that velocities comes directly from the gravitational potential (Eq. \ref{eq:radial_velocity}). The redshift-space power spectrum is then:
      \begin{equation}
     	  \label{eq:P_LN_s}
    	  P_{\rm LN}^s(k, \mu) = P_{\rm LN}(k) + f^2 \mu^4 P_{\rm G}(k) + 2 b f \mu^2 P_{\rm G}(k)
      \end{equation}
      where we used Eq. \ref{eq:xc_gaussian_lognormal} when computing the last term. 
      Similarly to the 2D clustering, we added a smoothing associated with the finite grid with $n_{\rm eff} = 1$. 
      We also added an extra smoothing to correct for the binning of the correlation function measurement.

      Differences on small scales between the measured and the predicted quadrupoles are due to inaccuracies in our RSD modelling, in particular higher-order terms ignored in Eq. \ref{eq:densityinredshift}.
      We discuss the role of these higher-order terms in Appendix \ref{app:higher_order_rsd}, and show that they are indeed the cause of this disagreement.

  \subsection{Performance at scale}\label{ssec:res.perf}
  
    In the context of existing and next-generation cosmological experiments, \colore{} should be able to generate mock observations covering large volumes ($z\lesssim3$), with reasonable resolution ($\Delta x={\cal O}(1)\,{\rm Mpc}$) for a wide range of observables (galaxy positions, shear, CMB lensing\footnote{Note that \colore{} can provide the contribution to the CMB lensing convergence up to the highest redshift covered by the simulation box. Contributions from higher redshifts can then be added as a correlated Gaussian random field.}, intensity maps, etc.). This section quantifies the feasibility of these simulations.

    \subsubsection*{Large-volume multi-tracer simulations}\label{sssec:res.perf.ubersim}
      \begin{figure}
        \centering
        \includegraphics[width=0.9\textwidth]{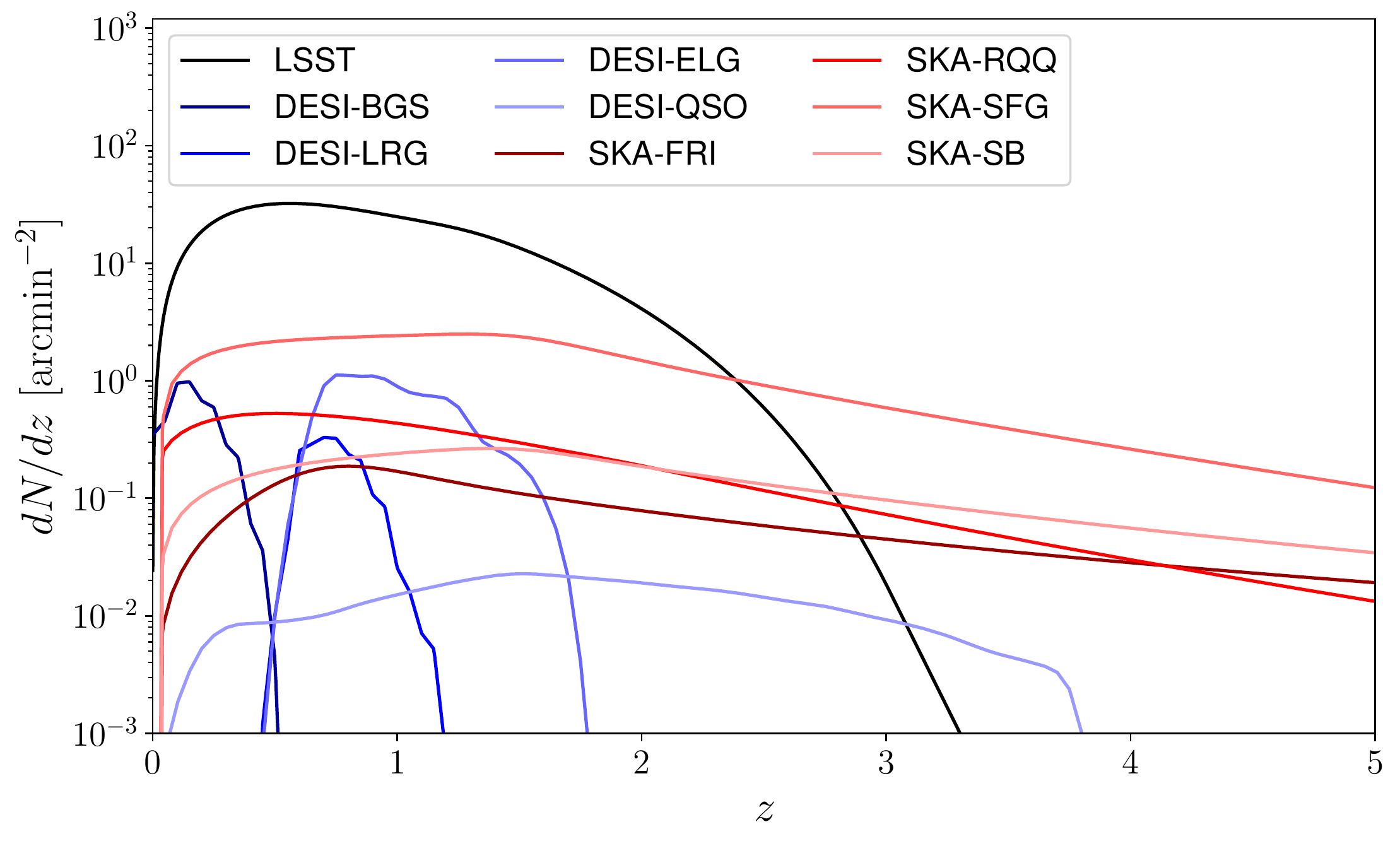}
        \caption{Redshift distribution of the different tracers simulated in our two flagship simulations. These include the LSST gold sample (black), the 4 DESI samples (blue) and 4 different radio continuum samples observable by SKA (red).}
        \label{fig:nz_uber_sim}
      \end{figure}
\begin{figure}
        \centering
        \includegraphics[width=0.99\textwidth]{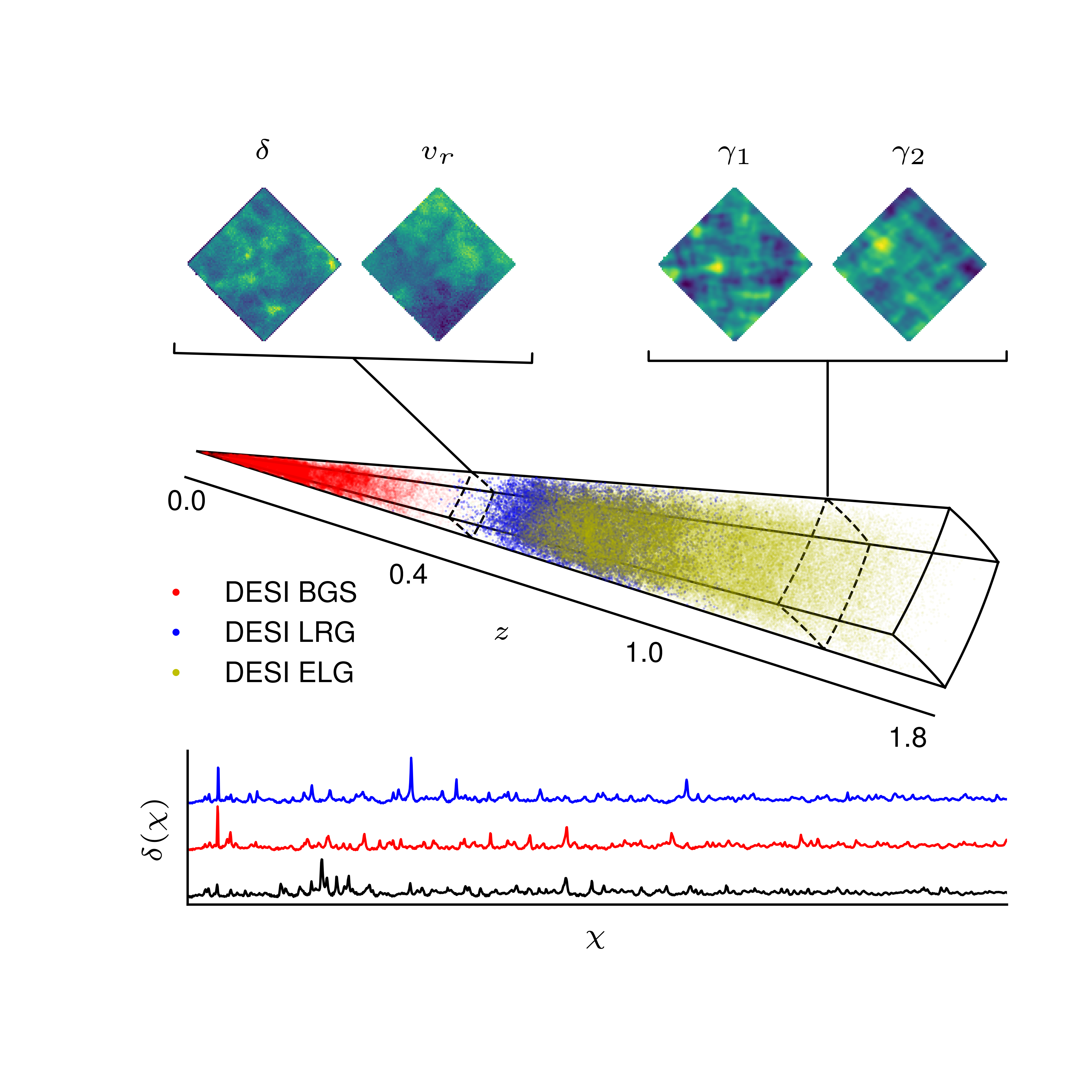}
        \caption{Visual description of the multi-tracer products that can be simulated with \colore{}. The upper plot shows one of the beams used internally by \colore{} for domain decomposition, with the redshift and angular coordinates of 3 different DESI samples, and maps of the density, radial velocity, and lensing shear constructed from sources at two different redshifts. The lower plot shows the density skewers calculated for three arbitrary DESI quasars contained in the same beam, as a function of comoving distance.}\label{fig:multitrace}
      \end{figure}

      As an example of the type of mocks needed for Stage-IV surveys, we have used \colore{} to generate two simulations containing cosmological probes for three major experiments:
      \begin{itemize}
        \item {\bf DESI.} We simulate the Bright Galaxy Survey (BGS), Large Red Galaxies (LRG), Emission Line Galaxies (ELG), and quasi-stellar object (QSO) targets (including Ly-$\alpha$ skewers for the QSO sample) using the nominal $N(z)$, and bias functions from the DESI white paper~\citep{DESI_whitepaper}. The resulting catalog contains $\sim3.4\times10^7$ BGSs, $\sim1.5\times10^7$ LRGs, $\sim10^8$ ELGs, and $\sim6.5\times10^6$ QSOs over the full celestial sphere.
        \item {\bf LSST.} We simulate a sample similar to the LSST "Gold" sample ($i\lesssim25.3$, $\bar{n}\sim 40$ galaxies/arcmin$^2$), resulting in $\sim6$ billion sources across the full sky. We follow the redshift distribution of the DESC Science Requirements Document~\citep{DESC-SRD}, and assume a linear bias with redshift dependence $b(z)=0.95/D(z)$~\citep{DESC-SRD, Nicola2020}. The lensing shear, convergence and displacement is calculated for all sources using the fast lensing scheme described in Section \ref{sssec:meth.tr.src}.
        \item {\bf SKA.} We simulate a radio continuum catalog comprised of 4 different radio galaxy samples: FRI radio-loud AGNs, radio-quiet AGNs (RQQs), normal star-forming galaxies (SFGs), and starbursts (SBs). For this we follow the models for the redshift distributions and linear bias described in \cite{2008MNRAS.388.1335W}. The resulting samples contain $\sim4.5\times10^7$ FRIs, $\sim1.3\times10^8$ RQQs, $\sim8\times10^8$ SFGs, and $\sim8\times10^7$ SBs over the full sky. In addition to this, we generate simulated HI intensity mapping observations for 490 frequency bands covering the range $473\,{\rm MHz}<\nu<947\,{\rm MHz}$ (corresponding to redshifts $0.5<z<2$). The maps were generated with an angular resolution $N_{\rm side}=256$, corresponding to pixels about 4 times smaller than the SKA primary beam in single-dish mode at the highest frequency ($\theta_{\rm FWHM}\sim1.2^\circ$).
      \end{itemize}
      In addition to these, to showcase the ability of \colore{} to generate CMB lensing observations, we generate a convergence map at resolution $N_{\rm side}=1024$ at $z=3$, caused by the same matter density field that serves as seed for the tracers listed above. The redshift distributions of the different galaxy samples simulated are shown in Fig. \ref{fig:nz_uber_sim}. For illustrative purposes, Fig. \ref{fig:multitrace} shows the 3-dimensional representation of some of the quantities simulated by \colore{} in one of the beams used by the code.

      The simulations were generated with a $\Lambda$CDM model compatible with the best-fit {\sl Planck} cosmological parameters~\citep{Planck18VI}. The boxes span the redshift range $(0 < z < 3)$, with $N_{\rm grid} = 4096$, resulting in a spatial resolution of $\approx 2$ Mpc/$h$.
      \begin{figure}
        \centering
        \includegraphics[width=0.99\textwidth]{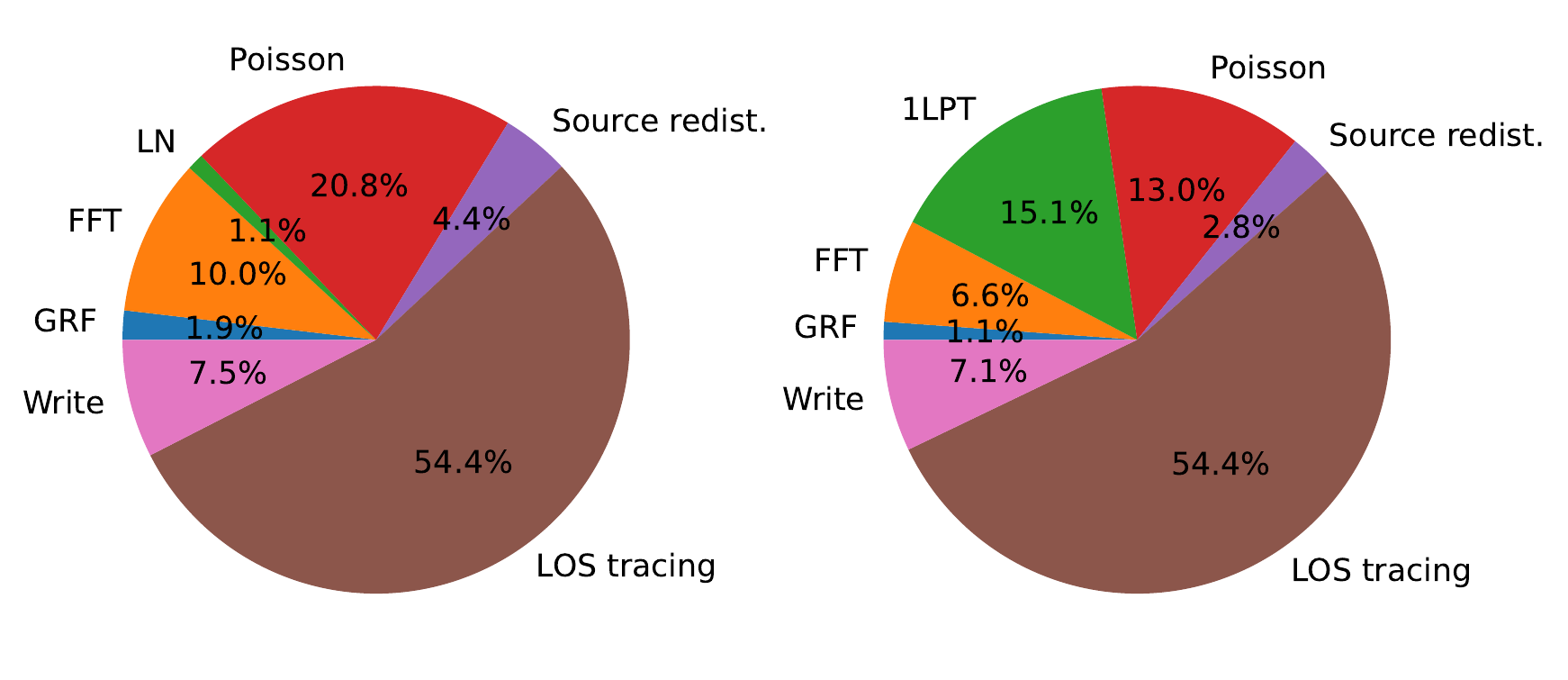}
        \caption{Fraction of the total run time taken up by different stages of a typical \colore{} simulation. The stages shown are, the generation of the initial Gaussian random fields in Fourier space ("GRF"), their transformation to real space ("FFT"), the structure formation model leading to a positive-definite matter overdensity ("LN" or "1LPT" for the lognormal and first-order LPT simulations), the generation of source catalogs via Poisson sampling ("Poisson"), the redistribution of these sources across different nodes before any line-of-sight calculations ("Source redist."), the calculation of all relevant line-of-sight quantities ("LOS tracing"), and the output of all final products to disk ("Write"). Note that the LOS-tracing stage is more sensitive to inter-node communication than other stages, and thus it takes up the same fraction of the total compute time in both simulations in spite of the additional time taken by the 1LPT stage, given the larger number of MPI nodes needed for that simulation.}
        \label{fig:pies}
      \end{figure}
      \begin{table}
        \begin{center}
          \begin{tabular}{|l|c|c|c|c|c|c|c|}
          \hline\hline
          Tracer & $\Delta z_{\rm RSD}$ & $\boldsymbol{\alpha}$ & $\boldsymbol{\gamma}$ & $\kappa$ & $\delta(z),v_r(z)$ & Memory (GB) & Disk (GB) \\
          \hline
          LSST     & \cmark & \cmark & \cmark & \cmark & \xmark & 308 & 155 \\
          DESI-BGS & \cmark & \xmark & \xmark & \xmark & \xmark & 1.7 & 0.6 \\
          DESI-LRG & \cmark & \xmark & \xmark & \xmark & \xmark & 0.8 & 0.3 \\
          DESI-ELG & \cmark & \xmark & \xmark & \xmark & \xmark & 5.2 & 1.9 \\
          DESI-QSO & \cmark & \xmark & \xmark & \xmark & \cmark & 99 & 99 \\
          SKA-FRI  & \cmark & \xmark & \xmark & \xmark & \xmark & 2.3 & 0.8 \\
          SKA-RQQ  & \cmark & \xmark & \xmark & \xmark & \xmark & 6.8 & 2.5 \\
          SKA-SFG  & \cmark & \xmark & \xmark & \xmark & \xmark & 39 & 15 \\
          SKA-SB   & \cmark & \xmark & \xmark & \xmark & \xmark & 4.2 & 1.5 \\
          SKA-21cm & \cmark & N.A. & N.A. & N.A. & N.A. & 2.8 & 1.4 \\
          Convergence map & N.A. & N.A. & N.A. & \cmark & N.A. & 0.1 & 0.05 \\
          \hline
          $\delta$, $\phi_N$ grids & N.A. & N.A. & N.A. & N.A. & N.A. & 768 & N.A. \\
          $\disp_{\rm 1LPT}$ grids & N.A. & N.A. & N.A. & N.A. & N.A. & 1229 & N.A. \\
          \hline
          \hline
          \end{tabular}
        \end{center}
        \caption{Simulation products generated by \colore{}. The first 11 rows show the different tracers generated for the large-volume simulations described in the text. For each tracer we show the physical quantities simulated (RSDs, lensing displacements, shear, convergence, and density/velocity skewers), as well as the memory and disk space taken. The last two rows display the memory requirements for the different Cartesian grids stored for lognormal and 1LPT simulations. Although the memory requirements are dominated by these cartesian grids (particularly for LPT simulations), the simulated tracers can take up a non-negligible fraction of the available memory. This is patent for the LSST sample, given its size, and the need to store lensing information, and for the DESI quasar sample, since we save a full density/velocity skewer for each source.}\label{tab:reqs}
      \end{table}

      \subsubsection*{Run time, memory usage, and fast lensing}
      Both simulations were initialised with the same random seed, but using different structure formation models, lognormal (LN) and first-order LPT (1LPT) respectively. Both simulations were run at NERSC~\footnote{\url{https://nersc.gov}}. The LN simulation was generated using 40 MPI tasks, distributed across 20 Cori-haswell nodes, using 16 OMP threads per task. This simulation ran in 1.15 hours, using approximately 730 CPU-hours. The 1LPT simulation required a larger number of nodes, given the additional memory needed to allocate the three more Cartesian grids mentioned in Section~\ref{sssec:meth.box.lpt}. This simulation was run using 72 MPI tasks distributed across 36 Cori-haswell nodes, using 16 OMP threads per task. The simulation ran in 0.93 hours using a total of 1,075 CPU-hours. A suite of 1000 such simulations could therefore be run using $\sim1$ million CPU-hours. 

      Table \ref{tab:reqs} lists the memory and disk requirements associated with each of these tracers\footnote{Note that, naively, we have simulated all galaxy tracers as disjoint samples when, in reality, there would be significant overlap between them (e.g. between SKA SFGs and the LSST sample). A more realistic setting should therefore account for these overlaps when generating the different galaxy catalogs.}. Although the most memory-demanding task is the generation of the 3D density and Newtonian potential fields, or the Lagrangian displacement components if using LPT, the final data products also lead to a non-negligible memory requirement, particularly in the case of high-density samples such as LSST, or for a large number of density/velocity skewers ($\sim27\%$ and $\sim4\%$ respectively in the case of the lognormal simulation).

      Figure \ref{fig:pies} shows the time taken by the different stages in both runs. In both cases the slowest stage is the collection of line-of-sight information, such as the density and velocity skewers stored for all DESI QSOs, or the lensing information associated with the LSST sources. Given the large number of sources in the LSST sample, this stage would completely dominate the run time if we had not used the ``fast lensing'' method described in Section \ref{sssec:meth.tr.src}.

      To quantify this, we ran two additional identical simulations with a total of $5.9$ billion galaxies, emulating a blue galaxy population of LSST 10-year depth. Both simulations were generated using 32 MPI tasks distributed across 16 Cori Haswell nodes at NERSC. Each MPI task had 16 OMP threads. The simulation using the fast lensing scheme took $\sim 35$ minutes to run (a total of $173$ CPU-hours), while the one integrating along each galaxy's line of sight took $\sim 36.3$ hours (i.e. $\sim 18,600$ CPU-hours), more than 60 times longer. The additional time was completely taken by the line of sight calculations. The fast-lensing scheme thus allows for a factor $\sim60$ speed-up and, in the current implementation of \colore{}, is absolutely necessary for simulations with billions of sources.
      \begin{figure}
	    \centering
    	\includegraphics[width=0.9\textwidth]{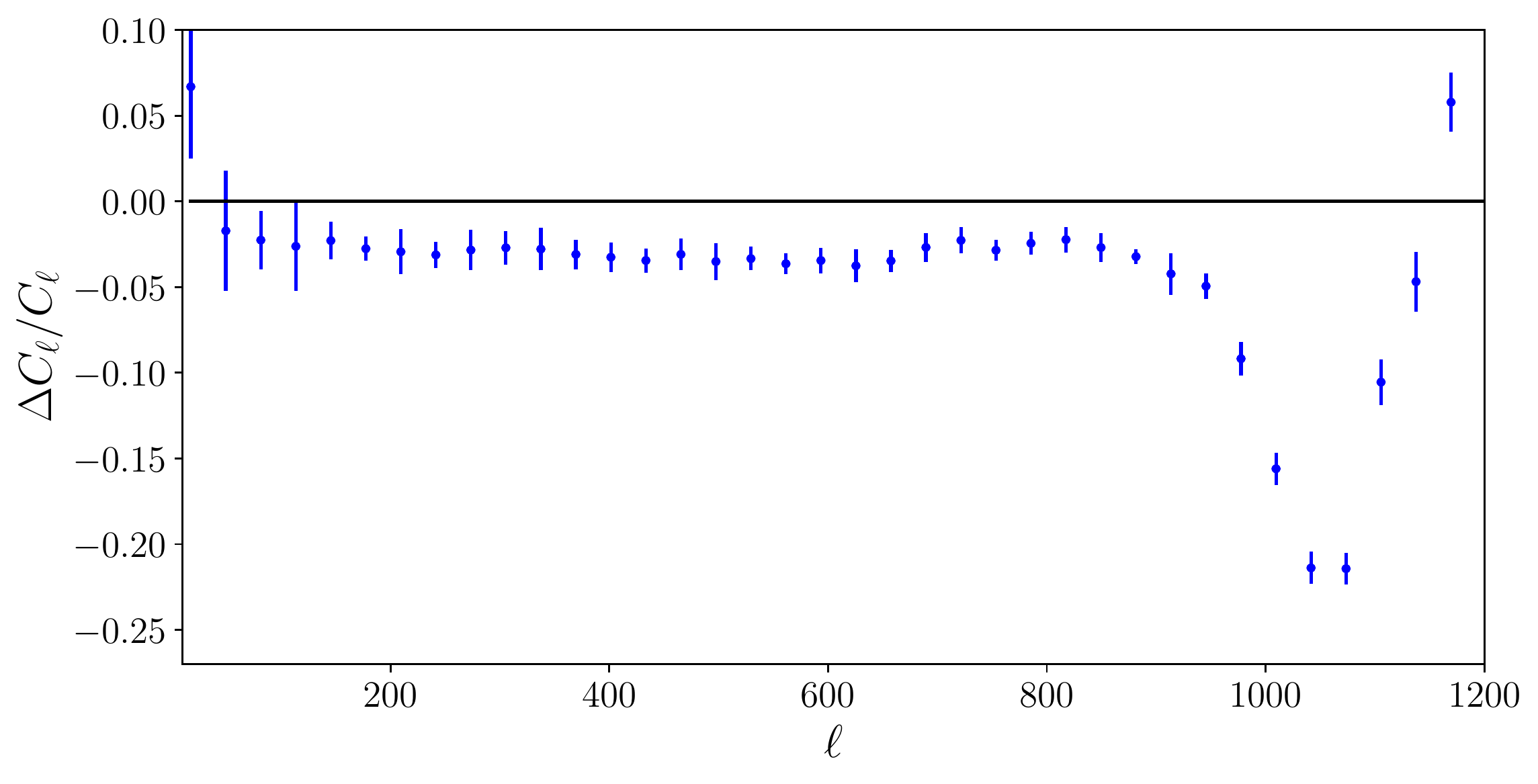}
	    \caption{Relative difference between the shear power spectrum of a \colore{} simulation run using the "fast lensing" scheme, and a simulation run without this approximation. The effects caused by the various interpolations carried out as part of the fast scheme should be carefully modelled if an accurate theoretical description of the \colore{} output is required.}\label{fig:cl_fastlens}
      \end{figure}

      The price to pay for this speed-up is the additional complexity associated with the interpolation operations involved in the fast scheme (interpolation from the cartesian grid to the fixed pixel lines-of-sight, and from those to the galaxy positions). Fig. \ref{fig:cl_fastlens} shows the relative difference between the cosmic shear power spectrum computed from a simulation run with the fast lensing scheme, and one run without this approximation. The effects of the fast lensing approximation are a decrease in the final lensing amplitude, and a gradual loss of power at higher multipoles, on scales comparable with the size of the adaptive pixels used in this approximation. These effects would need to be accurately characterized if the application of the \colore{} realizations requires an accurate theoretical prediction of two-point statistics involving weak lensing observables. The accuracy/speed-up trade-off can be controlled by the user by changing the number of shells in which to compute the lensing information, as well as the size of the adaptive pixels in relation with the Cartesian cell size.

\section{Conclusions}\label{sec:conclusions}
  In this paper we have introduced \texttt{CoLoRe}, a public code to efficiently generate synthetic realisations of multiple cosmological surveys. We started in Section 2 by describing the overall structure of the code, and the different methods to simulate the density field. We have presented the available tracers in \texttt{CoLoRe}, and how to add new ones using its highly modular structure. We concluded this section discussing the accurate predictions available when working with the lognormal model of structure formation. 

  In Section 3 we presented the validation results of some of the key summary statistics from the simulated maps, using a large set of \texttt{CoLoRe} boxes. We showed that the measured angular power spectra from simulated photometric surveys agree with the theoretical predictions up to $\ell=1000$. This is true for galaxy correlations, galaxy-shear cross-correlations and several lensing statistics (shear, convergence, displacements). The 3D clustering in simulated spectroscopic galaxy surveys was also validated in the absence of redshift space distortions, where the lognormal model can be accurately predicted on all scales. As discussed in Appendix \ref{app:higher_order_rsd} we do not have a good model for the small-scales multipoles in redshift space, but the agreement is very good on large, linear scales.

  We have discussed the performance of \texttt{CoLoRe} at scale by presenting two joint simulations of DESI, LSST and SKA. These large boxes cover the whole comoving volume out to $z=3$, with a resolution of $ \approx 2 \rm{Mpc}/h$, and include spectroscopic and photometric galaxies, lensing, intensity mapping and radio galaxies. The more realistic simulation, using Lagrangian perturbation theory, only used about 1,000 CPU-hours. Simulating hundreds of these boxes is entirely feasible, and can be used to characterise systematic effects in multi-experiment analyses, or estimate cross-survey covariances.

  Finally, we discussed the differences between the two options used in \texttt{CoLoRe} to compute weak lensing variables from source galaxies. The fast lensing implementation provides a factor of $\sim$60 speed-up for an LSST-like sample, while maintaining an accuracy better than a 2-3\% bias in the amplitude of the shear power spectrum on large scales ($\ell < 1000$).
  
There are several features that could be added to \texttt{CoLoRe} without major changes in the code structure:
\begin{itemize}
  \item {\bf Better structure formation:} Currently \texttt{CoLoRe} can simulate the growth of structure using a lognormal model or Lagrangian Perturbation Theory (LPT) computed at first or second order. Future versions of the code could add new modules to use more complex models of the growth of structure, such as COmoving Lagrangian Acceleration (COLA, \citep{2013JCAP...06..036T} as in \citep{aizard_icecola_lensing_2017}, where it was used to generate weak lensing maps and halo catalogues in the lightcone) or Fast Particle-Mesh (FastPM, \citep{2016MNRAS.463.2273F}) algorithms. The potential additional compute time and memory requirements associated with these methods should be weighed against the need for more accurate clustering statistics of a given application.
  \item {\bf Non-linear RSD:} Minor modifications of the code could improve the level of realism of the RSDs, by sourcing the velocities from the computed LPT fields. This change could have a significant impact on the 3D clustering of galaxies intermediate scales. However, CoLoRe does not currently simulate virialized objects, and therefore we are not able to properly capture non-linear peculiar velocities or Fingers of God. Random virial motions (or redshift errors) can be added in post-processing, by assigning random shifts to the galaxy redshifts, but a more realistic approach of non-linear RSD is beyond the scope of this paper.
  \item {\bf Halos:} The models currently used by \colore{} to generate different tracer observations directly connect the latter with the underlying smooth density field. The complexity and fidelity of these simulations could be improved if this density field was endowed with a halo catalog. This could be done directly at the level of the linear density field using Press-Schechter-inspired methods (as in e.g. \cite{2010MNRAS.406.2421S}), or from the LPT displacement field using a modified friends-of-friends search \citep{2013MNRAS.428.1036M}, or peak-patch methods \citep{2019MNRAS.483.2236S,2020JCAP...10..012S}. A halo catalog would allow us to improve the fidelity of the resulting density field on small scales by including the expected density profile of these halos, and would make it possible to use the halo model to generate simulated observations of additional tracers (e.g. thermal or kinematic Sunyaev-Zel'dovich effects, halo-occupation distributions for better galaxy catalogs etc.).
  \item {\bf Small areas and flat-skies:} The current version of \texttt{CoLoRe} simulates the whole Universe to a given redshift, with the observer placed in the center of a large box.
  Users interested in simulating surveys with relatively small areas might prefer to place the observer in one side of rectangular box and simulate only a (literal) light-cone. For sufficiently small sky areas, these simulations could be made faster making use of the flat-sky approximation. Both of these features should be easy to implement in \colore{}.
\end{itemize}

The addition of these features should not significantly impact the performance of \colore{} reported here. By making our code public we want to encourage the different collaborations preparing the next generation of large cosmological surveys to use \texttt{CoLoRe} to efficiently generate realistic synthetic simulations of their datasets to be used in multi-survey analyses.

\acknowledgments
We would like to thank James Farr, David Kirkby, Stephane Plaszczynski, and An\^{z}e Slosar  for contributions to the \colore{} repository and for useful discussions in early stages of this project. We would also like to thanks Hanyu Zhang for his assistance with the measurement of the 3D correlations using \texttt{corrfunc}. CRP is partially supported by the Spanish Ministry of Science and Innovation (MICINN) under grants PGC-2018-094773-B-C31 and SEV-2016-0588. DA is supported by the Science and Technology Facilities Council through an Ernest Rutherford Fellowship, grant reference ST/P004474. AFR is supported by MICINN with a Ram\'on y Cajal contract (RYC-2018-025210). IFAE is partially funded by the CERCA program of the Generalitat de Catalunya.

We made extensive use of computational resources at the University of Oxford Department of Physics, funded by the John Fell Oxford University Press Research Fund. We also used resources of the National Energy Research Scientific Computing Center (NERSC), a U.S. Department of Energy Office of Science User Facility located at Lawrence Berkeley National Laboratory, operated under Contract No. DE-AC02-05CH11231. This manuscript has been authored by Fermi Research Alliance, LLC under Contract No. DE-AC02-07CH11359 with the U.S. Department of Energy, Office of Science, Office of High Energy Physics.

We made extensive use of the {\tt numpy} \citep{oliphant2006guide, van2011numpy}, {\tt scipy} \citep{2020SciPy-NMeth}, {\tt astropy} \citep{astropy:2013, astropy:2018}, {\tt healpy} \citep{Zonca2019}, {\tt NaMaster} \citep{2019MNRAS.484.4127A}, {\tt corrfunc} \citep{corrfunc}, {\tt matplotlib} \citep{Hunter:2007} and {\tt lmfit} \citep{newville_matthew_2014_11813} python packages. 

\bibliography{bibliography}

\appendix

\section{Higher-order terms in the modelling of redshift-space distortions}\label{app:higher_order_rsd}
\newcommand{\dLN}{\delta_{\rm LN}}
\newcommand{\dLNs}{\delta^s_{\rm LN}}
\newcommand{\dLNns}{\delta^{-s}_{\rm LN}}
\newcommand{\vx}{\mathbf{x}}
\newcommand{\vk}{\mathbf{k}}
\newcommand{\PG}{P_{\rm G}}
\newcommand{\PLN}{P_{\rm LN}}

  The redshift-space galaxy overdensity, $\dLNs=\dLN(\mathbf{s})$, is related to its real-space equivalent $\dLN(\vx)$ and the normalized gradient of line-of sight velocities $\eta = - \partial_z v_z / H(z)$ via:
  \begin{equation}
    \label{eq:delta_LN_s_ext}
    1 + \dLNs = \frac{1+\dLN}{1-\eta}
  \end{equation}
  Assuming Gaussian RSDs are small, we can expand this in powers of $\eta$,
  \begin{equation}
    \dLNs = \dLN + \eta + \epsilon ~,
  \end{equation}
  where $\epsilon (\vx) \equiv \dLN (\vx) ~ \eta(\vx)$. This term would be order-2 assuming $\dLN$ is small, but we do not make this approximation here. Ignoring this term one would recover Eq. \ref{eq:densityinredshift}, used for the theoretical predictions in Section \ref{sssec:res.val.3d}. If we keep this extra term, however, the model for the redshift-space galaxy power spectrum will have new contributions with respect to the model described in Eq. \ref{eq:P_LN_s}:
  \begin{align}
    \label{eq:P_LN_s_ext}
    \PLN^s (k, \mu) = & ~ \left( \Delta k \right)^3 \langle |\dLNs (\vk)|^2\rangle           \nonumber \\
        = & ~ \PLN (k) + f^2 \mu^4 \PG (k) + 2 b f \mu^2 \PG(k)       \\
        & ~ + \left( \Delta k \right)^3 \left[ 2 \langle \dLN (\vk) ~ \epsilon (\vk) \rangle
            + 2 \langle \eta (\vk) ~ \epsilon (\vk) \rangle
            + \langle \epsilon (\vk) ~ \epsilon (\vk) \rangle \right] ~.        \nonumber
  \end{align}

\newcommand{\dA}{\delta_A}
\newcommand{\dAs}{\delta_A^s}
\newcommand{\dB}{\delta_B}
\newcommand{\dBs}{\delta_B^s}

\begin{figure}[h]
  \centering
  \includegraphics[width=\textwidth]{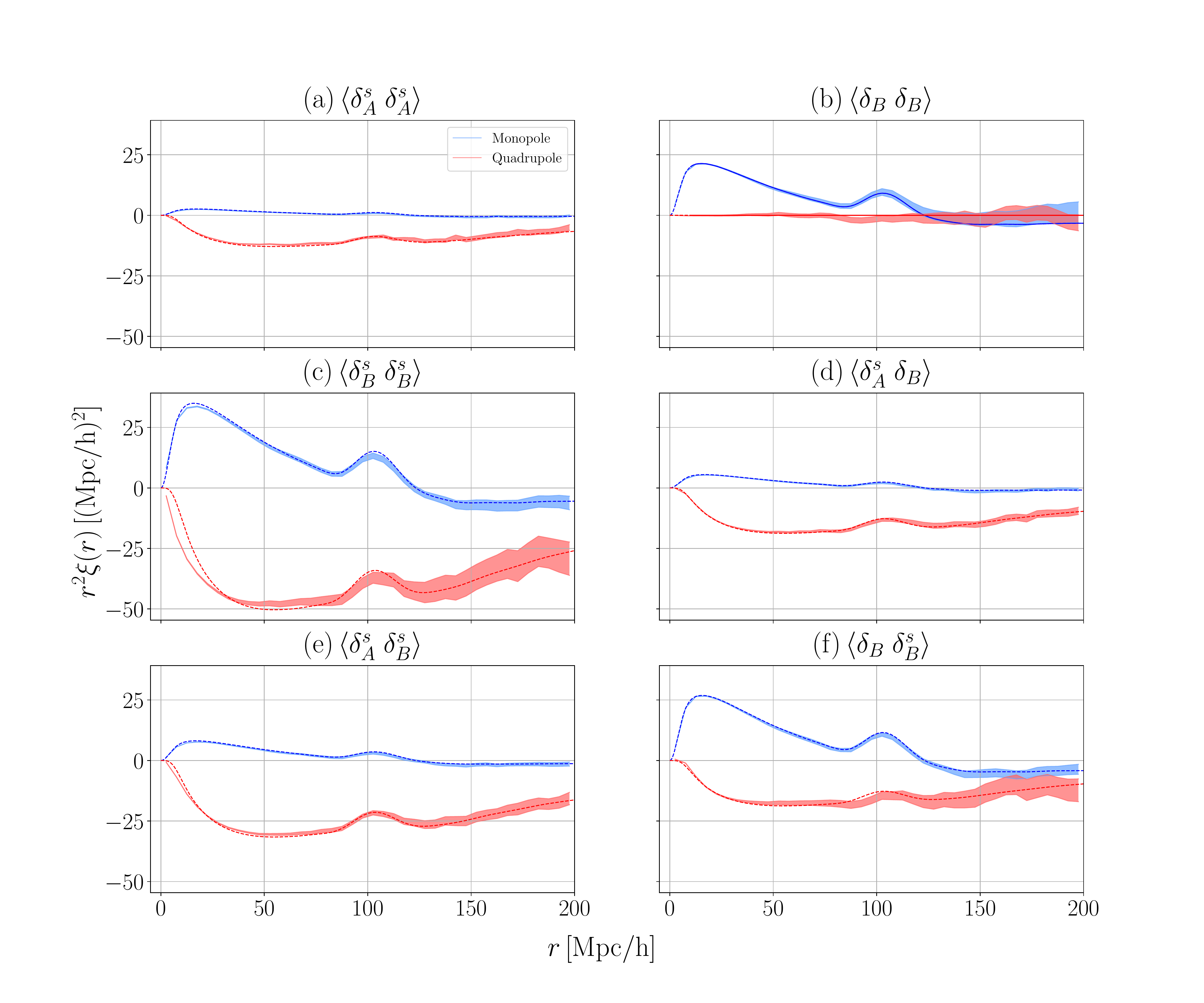}
  \caption{Measurements and predictions for the different cross-correlations discussed in \ref{app:higher_order_rsd} for a single simulation with two special tracers ($b_A=0.001$ and $b_B=2$).
  The predictions are from the simpler model discussed in \ref{sssec:res.val.3d} and are missing those terms involving $\epsilon(\vx)=\dLN(\vx) \eta(\vx)$.
  This explains the disagreement seen on the small-scales quadrupole in panel (c), it also shows a small disagreement in panels (e) and (f).
  }
  \label{fig:3d_test}
\end{figure}

  Even though it is possible to compute analytical predictions for the correlation function including these new $\epsilon$ terms (as convolutions of $\dLN$ and $\eta$ in Fourier space), we leave this for future work. Here we only quantify each of the terms by looking at cross-correlations of test galaxy samples from a custom \colore{} simulation designed for this study. This simulation was similar to those described in \ref{sssec:res.val.sim} but with two spectroscopic galaxy samples: a first sample $\dA$ with an extremely low clustering amplitude ($b_A=0.001$), and a second sample $\dB$ with a redshift-independent bias of $b_B=2$.

  While the clustering of sample $A$ in real space is negligible, its redshift-space power spectrum can be modeled as $\langle \dAs (\vk) \dAs (\vk) \rangle \propto f^2 \mu^4 \PG(k)$ (up to a normalisation factor $(\Delta k)^3$). This can be clearly seen in the (a) panel of Figure \ref{fig:3d_test}, and validates the simulation of RSDs in \texttt{CoLoRe}. The clustering of sample $B$ in real (redshift) space is shown in the (b) ((c)) panels of the same figure, and are similar to Fig. \ref{fig:3d_clust} discussed in \ref{sssec:res.val.3d}. While its real space clustering is well described by the lognormal model $\langle \dB (\vk) \dB (\vk) \rangle \propto \PLN(k)$, the small-scales quadrupole of its redshift-space equivalent can not be described by the simple model of Eq. \ref{eq:P_LN_s}. We are missing the contributions from the terms $\langle \dLN ~ \epsilon \rangle$, $\langle \eta ~ \epsilon \rangle$ and $\langle \epsilon ~ \epsilon \rangle$ introduced in Eq. \ref{eq:P_LN_s_ext}.

  In panel (d) we show the cross-correlation of $\dAs$ and $\dB$, compared to its prediction $\langle \dAs (\vk) \dB (\vk) \rangle \propto b_B f \mu^2 \PG(k)$. Because there is no $\epsilon$ term involved in this cross-correlation, the model describes the measurement very well on all scales. In panel (e) we cross-correlate $\dAs$ with $\dBs$ instead, and compare it to a theoretical prediction that includes the two Kaiser terms ($b_B f \mu^2 \PG(k) + f^2 \mu^4 \PG(k)$) but is missing an extra term $\langle \eta ~ \epsilon \rangle$. The minor disagreement of the quadrupole on small scales allows us to estimate the magnitude and sign of the missing term. Finally, in (f) we show the cross-correlations of $\dB$ and $\dBs$, i.e., the cross-correlation of the same B sample in real and in redshift space. We plot its prediction from the simpler model from section \ref{sssec:res.val.3d}, including the terms $\PLN(k) + b_B f \mu^2 \PG(k)$. Again, following Eq. \ref{eq:P_LN_s_ext} this cross-correlation should include an extra term $\langle \dLN ~ \epsilon \rangle$ that explains the small disagreement of the quadrupole on small scales. 

\begin{figure}[h]
  \centering
  \includegraphics[width=0.95\textwidth]{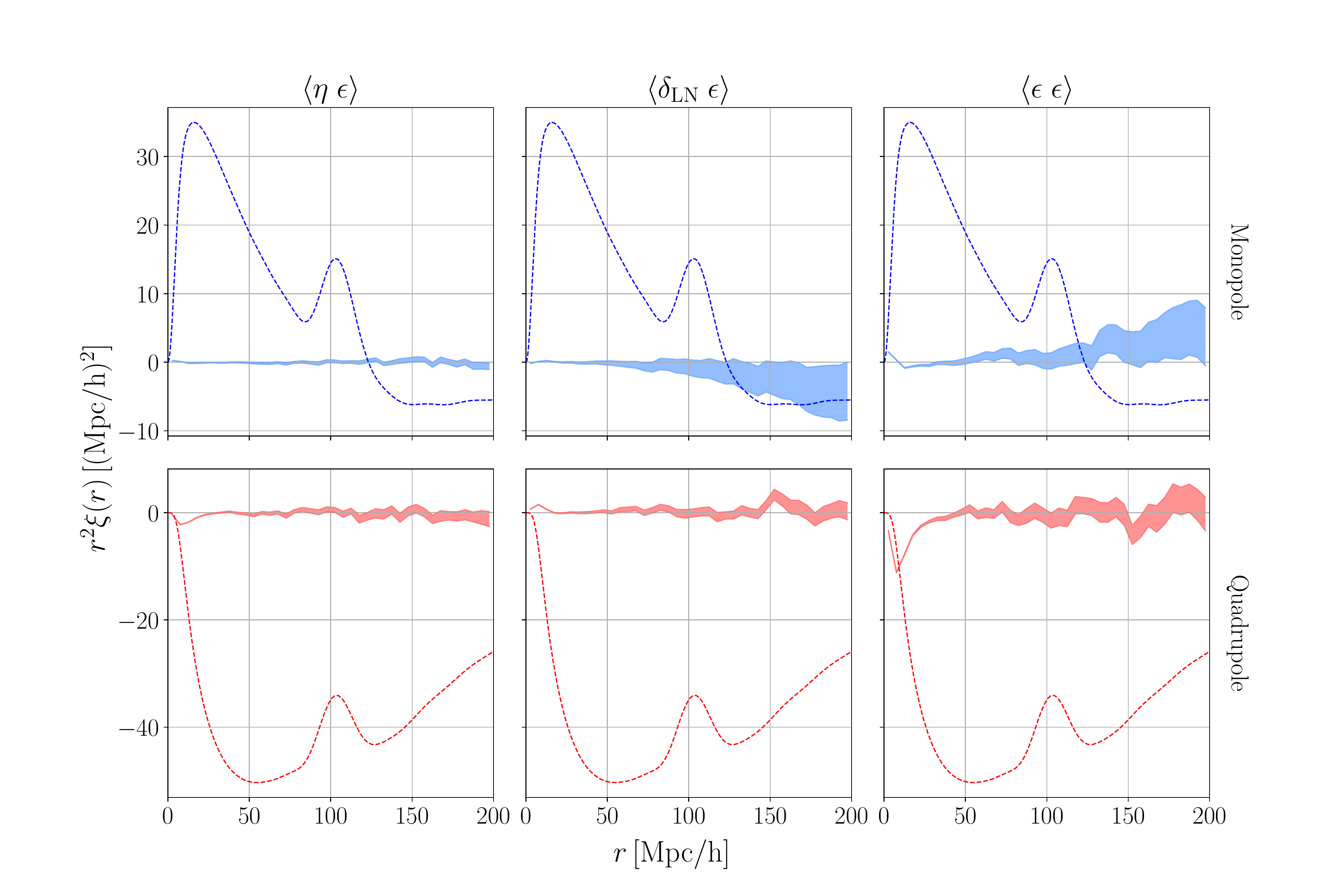}
  \caption{Contributions to the monopole (top panel/blue lines and bands) and the quadrupole (bottom panel/red lines and bands) from the terms in Eqs. \ref{eq:app1_term_11}-\ref{eq:app1_term_13} using the same simulation as in Figure \ref{fig:3d_test}. The shaded bands show the error in the measurement of each term from the scatter between 48 healpixels of $N_{\rm {side}}=2$. The dashed lines show the full model prediction as in section \ref{sssec:res.val.3d}, where the extra terms are not included. 
  Particularly the term $\langle \epsilon ~ \epsilon \rangle$ (right panel) has an important impact on the small-scales quadrupole.
  }
  \label{fig:3d_ext}
\end{figure}

Following Eq. \ref{eq:delta_LN_s_ext} it is clear that $\epsilon = \dBs - \dB - \dAs$.
Therefore, by combining the different cross-correlations discussed above we can now isolate each of the new terms in \ref{eq:P_LN_s_ext}:
\begin{align}
    \langle \dLN ~ \epsilon \rangle 
        = & ~ \langle \dB ~ \dBs \rangle - \langle \dB ~ \dB \rangle - \langle \dB ~ \dAs \rangle    \label{eq:app1_term_11}\\
    \langle \eta ~ \epsilon \rangle 
        = & ~ \langle \dAs ~ \dBs \rangle - \langle \dAs ~ \dB \rangle - \langle \dAs ~ \dB \rangle  \label{eq:app1_term_12}\\
    \langle \epsilon ~ \epsilon \rangle
        = & \langle \dBs ~ \dBs \rangle + \langle \dB ~ \dB \rangle + \langle \dAs ~ \dAs \rangle      
        + 2 \langle \dAs ~ \dB \rangle - 2 \langle \dAs ~ \dBs \rangle - 2 \langle \dB ~ \dBs \rangle ~. \label{eq:app1_term_13}     
\end{align} 
The multipoles corresponding to these three combinations of correlations are shown in Figure \ref{fig:3d_ext}, together with the model prediction from equation \ref{eq:P_LN_s}.
One can see that the largest correction to the model should come from the $\langle \epsilon ~ \epsilon \rangle$ term, while the other two terms are small and partially cancel each other.

\end{document}